\documentclass[aps,prb,twocolumn]{revtex4}
\usepackage{amsmath,epsfig,mathrsfs,graphicx,amssymb}
\usepackage{amsmath,epsfig,mathrsfs,graphicx,amssymb}

\def\be#1\ee{\begin{equation}#1\end{equation}}

\newcommand{\ba}{\begin{eqnarray} }
\newcommand{\ea}{\end{eqnarray} }
\begin{document}

\title{Models of mesoscopic time-resolved current detection}

\author{Adam Bednorz$^{1,2}$ and Wolfgang Belzig$^1$}
\affiliation{$^1$Fachbereich Physik, Universit{\"a}t Konstanz, D-78457 Konstanz, Germany\\
$^2$University of Warsaw, Ho\.za 69, PL-00681 Warsaw, Poland} 

\date{\today}

\begin{abstract}
  Quantum transport in mesoscopic conductors is essentially governed
  by the laws of quantum mechanics. One of the major open questions of
  quantum mechanics is what happens if non-commuting observables are
  measured simultaneously. Since current operators at different times
  do not commute, the high-frequency correlation functions of the
  current are realization of this fundamental quantum question.  We
  formulate this problem in the context of measurements of
  finite-frequency current cumulants in a general quantum point
  contact, which are the subject to ongoing experimental effort.  To
  this end, we present two models of detectors that correspond to a
  weak time-resolved measurement of the electronic current in a
  mesoscopic junction. In both cases, the backaction of the detector
  leads to observable corrections to the current correlations
  functions involving the so-called noise susceptibilities. As a
  result, we propose a reinterpretation of environmental
  corrections to the finite-frequency cumulants as inevitable effect
  resulting from basic quantum mechanical principles. Finally we make
  concrete predictions for the temperature-, voltage-, and
  frequency-dependence of the third cumulant, which could be verified
  directly using current experimental techniques. 


\end{abstract}
\maketitle

\section{Introduction}

Since the advent of mesoscopic physics, the quantum mechanical
properties of charge transport in conductors are under intensive
research. Most manifestations to date are concerned with properties
following from the quantum mechanical wave nature.
\cite{altshuler} More recently the quantum mechanics of the electron
spin has been investigated in electron transport. \cite{awschalom} On
the other hand, the aspect of measurement has attracted less attention
so far in quantum transport. This is surprising in view of it's
eminent importance for the fundamental difference to classical physics.
\cite{neumann}

The electronic transport through mesoscopic junctions is quite well
described by independent electrons. \cite{blanter} Due
to the Landauer formula, the conductance can be expressed in terms of
the single-particle scattering matrix and the level occupations of the
leads. The same scattering matrix can describe low-frequency current
noise. \cite{blanter} Experimentally, the scattering description has
been confirmed in the case of low frequency noise
\cite{glatt1} a while ago.

It is remarkable, that the same scattering matrix also describes
higher order correlation functions of current -- leading to the
so-called full counting statistics (FCS).\cite{fcs} These results have
been confirmed for the third cumulant of tunnel junctions only, so
far. \cite{reulet,reznikov} However, a proper account of the first
experiments showed that the measured third cumulant was governed by an
environmental contribution. \cite{been}

In the last years it has also been realized, that the measurement of
the quantum noise depends in an essential way on the detection
scheme. \cite{schoellkopf-yuliproc} To understand this, one considers
different Fourier transform of the current-current correlation
function $C(t,t^\prime)=\langle \hat I(t)\hat I(t^\prime)\rangle$. It
turns out that the unsymmetrized correlators $C(\omega)=\int d\tau
C(0,\tau) \exp(\pm i\omega \tau)$ are related to absorption or
emission of an energy quantum $\hbar\omega$, respectively, and can
therefore be measured in a corresponding detector. On the other hand,
a classical detector measures always the real combination
$C(\omega)+C(-\omega)$, corresponding to the anti-commutator of the
current operators. The high frequency noise has been also tested
experimentally. \cite{schoel,glatt2,gabel} The results are in full
agreement with the scattering theory of non-interacting quasiparticles
in mesoscopic conductors.

From the above it is clear that a time-resolved detection of the
quantum mechanical current in a quantum transport process has to
properly account for the quantum rules for measurements. This is not
so much an experimental question -- since experimentalists
automatically obey the rules of quantum mechanics -- rather, the proper
treatment of a quantum measurement represents a challenge to
theorists, who have to worry, which expectation values are actually
measurable. Hence, the extension of full counting statistics to
time-resolved (high-frequency) correlations requires first a proper
definition in terms of projective measurement.\cite{neumann}
Qualitatively, one can distinguish two extremes. A continuous
projective measurement would lead up in suppression of the dynamics of
the system -- a phenomenon known as quantum Zeno effect. \cite{zeno} To
avoid the Zeno effect it is necessary to include the detector's
degrees of freedom into the complete evolution. Some effects of the
detector backaction have been addressed already in the
literature.\cite{brag,jordan,ave,les08} However, instead of including
a specific detector's dynamics one can restrict the treatment to the
system only by replacing the projection by so-called Kraus operators.
\cite{kraus} These define a positive operator-valued measure
(POVM) \cite{povm} and by virtue of Naimark's theorem the two methods are
equivalent \cite{naim} as long as initially the detector and the
system are uncorrelated. \cite{haenggi}

Once we have established the detection model we can again express
current correlations of arbitrary order in terms of the
single-particle scattering matrix.  However, it turns out that a
simple evaluation of these averages quickly leads to very complicated
and cumbersome expressions.  Nevertheless, the demanding task
necessary to analyze the generating functional leads to interesting
analytic results for the frequency-dependent cumulants and the full
counting statistics.\cite{zaikin1,salo,heikkilae:07} Below, we will provide a
simplified framework, which allows to evaluate correlators of
arbitrary order in an efficient way.

An important recent development is that the third cumulant has been
measured at high frequency.\cite{regab} These quite remarkable results
show interesting quantum features at $eV=\hbar\omega$. However, most
of this feature can be explained by environmental effects due to a
series impedance. These result in a mixing of the third-order
correlator with the noise susceptibilities due to the
frequency-dependent backaction of an environmental impedance. After
fitting the spurious contribution, the resulting third cumulant of the
considered tunnel junction shows no frequency dispersion at all. This
is in perfect accordance with the theory, which predicts in fact a
dispersionless third cumulant in the case of a tunnel
junction. \cite{zaikin1} However, it is most likely just a question of
time until the third cumulant for a mesoscopic conductor with
non-opaque channel is measured. Hence, for these experiments it is
vital to distinguish the environmental contribution from the
quantum mechanical backaction of the detector.

Our findings can be summarized in two main achievements. On one hand,
most of the article is devoted to the development of concrete models of
mesoscopic detectors of high-frequency current-detection. The
interplay between the quantum mechanical projection and backaction
plays here a major role and we clarify below, how this can or cannot
be distinguished from the environmental effect. On the other hand, we
present an efficient method to calculate higher order correlation
functions of the current in mesoscopic junctions. This allows us to
present several results for the third cumulant and the noise
susceptibility in a unified and transparent manner. The method can be
the basis for an efficient evaluation of fourth and higher order
correlators, which is important for example in photon counting
statistics.\cite{schome,gabelli:04} 

In the main body of the article we are going to construct a model of a
projective time-resolved measurement of the current in a mesoscopic
conductor.  We present two possibilities. One is to define a
POVM, which corresponds to a weak measurement of the current in a
quantum point contact. This phenomenological and very simple approach
has the appeal that the competition between projection and weak
measurement is qualitatively reflected. However, a concrete physical
realization is difficult to find.  Another possibility is based on a
physical model of a detector, which we call \emph{quantum tape}. Our
detector will be a massless bosonic field described by charge density
and current operators, parametrized by a one-dimensional variable. It
is equivalent to the noninteracting Schwinger-Tomonaga-Luttinger model
of a one-dimensional fermionic field. \cite{lut} Equivalently the
detector can be thought of as a transmission line, which is described
by the same Hamiltonian. This detector model is actually a massless
version of a Josephson transmission line detector. \cite{ave} The
field can be decomposed into left and right going components with a
constant velocity.  In this way, the spatial coordinate of the
detector, which is coupled to the system only at one point in space,
corresponds effectively to a time-coordinate. The information
registered by the detector is moved away from the system due to
the internal dynamics of the detector.  Finally a spatially resolved
strong projection is applied to the detector at one instant of time,
and the result can be translated into the temporal current profile. The
projection can be made on one or both ends of the tape due to its
bidirectional information transfer. The results always depend on the
detection scheme. In the limit of weak coupling the detector's
contribution is only a large Gaussian offset noise, independent of the
measured system.\cite{ours} Accordingly, the influence of the
backaction of the detector onto the system is also weak but always
present.

The secondary goal of the present article is to develop a compact
method to calculate arbitrary cumulants of current operators in a
junction with energy-independent transmission. To this end, we set up
a current algebra, which is especially suited for non-equilibrium
scattering problems. This becomes necessary, when considering
different orderings of higher-order correlators at finite frequency. A
straightforward evaluation using the scattering states is possible,
but becomes in general very cumbersome. Introducing an algebra for
certain current operators of a quantum point contact greatly
simplifies this calculation and will therefore be useful also in other
context, hopefully. In the respective limits, our results are, of
course, in agreement with existing results.

The paper is organized as follows. We begin with the description of
the mesoscopic junction in Sec. II, together with an overview of
necessary mathematical tools.  In Sec. III we present the POVM
detection model and extend it to many detectors.  In this model, we
are able to derive also some higher cumulants of current in the limit
of weak coupling between the detector and the system. In certain
limits (weak transmission or reflection) one can find the complete FCS
generating functional.  Finally in Sec. IV, we present a strictly
projective measurement model -- the quantum tape, which gives results
similar to the POVM. Many useful but lengthy mathematical details are
moved to appendices.


\section{Mesoscopic Junction}

The junction is defined as constriction in a two dimensional electron
gas, which is narrow in the $y$-direction and relatively long in the  
$x$-direction.
We assume no electron-electron interactions and no magnetic field. The
Hamiltonian of the system reads 
\ba
&&\nonumber\hat{H}=\sum_\sigma\int dxdy\left[-\hat{\psi}^\dag_\sigma(x,y)\hbar^2\Delta\hat{\psi}_\sigma(x,y)/2m\right.\\
&&\left.+ eV_e
(x,y,t)\hat{\psi}^\dag_\sigma(x,y)
\hat{\psi}_\sigma(x,y)\right].\label{ham1}
\ea
The fermionic operators satisfy standard anticommutation relations
\ba
&&\{\hat{\psi}_{\sigma}(x,y),\hat{\psi}_{\sigma'}(x',y')\}=0,\nonumber\\
&&\{\hat{\psi}_{\sigma}(x,y),\hat{\psi}^\dag_{\sigma'}(x',y')\}=\delta_{\sigma\sigma'}\delta(x-x')
\delta(y-y').
\ea
The external potential splits into three parts
\begin{equation}
V_e(x,y,t)=V_x(x)+V_y(y)+V(x,t),
\end{equation}
where $V_x$ is the scattering potential, $V_y$ is the effect of the
constriction and $V$ is the time-dependent bias potential. The
scattering potential is assumed to be nonzero only in a small interval
around the center of the constriction while $V_y$ grows to infinity
with $|y|\to\infty$. We have the  total current operator in $x$
direction 
\ba
&&\hat{I}(x)=-\frac{i\hbar e}{2m}\int dy\\
&&
\sum_\sigma[\hat{\psi}_\sigma^\dag(x,y)\partial_x\hat{\psi}_\sigma(x,y)-
(\partial_x\hat{\psi}_\sigma^\dag(x,y))\hat{\psi}_\sigma(x,y)].\label{cur1}\nonumber
\ea

In absence of a bias voltage, the Hamiltonian can be diagonalized in
the space of scattering states described by fermionic operators
$\hat{\psi}_{k,\bar n}(x)$ with the channel quantum number including
the electron spin $\bar n=(n,\sigma)$.  Far from the region $V_x\neq
0$, we have
\ba
\nonumber
\hat{\psi}_{\sigma}(x,y) & = & \int \frac{dk}{\sqrt{2\pi}}
\psi_{n}(y)\left[(e^{ik x}+\tilde{r}_ne^{-ik x})\theta(-kx)\right.\\
&& \left.+ \tilde{t}_ne^{ik x}\theta(kx)\right]\hat{\psi}_{k,\bar n}
\label{sct}
\ea
Here we distinguish between the normalized transversal mode
wavefunction $\psi_n$ and the mode operator $\hat{\psi}_{k,\bar n}$, $k$ is the longitudinal wave vector, $\tilde{r}_n$ and
$\tilde{t}_n$ are reflection and transmission scattering amplitudes,
respectively. Due to unitarity, transmission $T_n=|\tilde{t}_n|^2$ and
reflection coefficient $R_n=|\tilde{r}_n|^2$ satisfy $R_n+T_n=1$.  We
assume that the junction is long enough to treat $k$ as a continuous
parameter. The transverse mode index $n$ is kept as discrete.  The
energy has the structure $ E=E_n+\hbar^2k^2/2m$, where $E_n$ is for
the transversal part.

\subsection{Expansion around Fermi level}

We are interested only in phenomena in a narrow part of the electron
band around Fermi level $E_F$.\cite{blanter} In particular, we assume
$k_BT,eV,\hbar|\omega| \ll E_F$.  We also neglect the exact structure
of wavefunctions in the region $V_x\neq 0$ and take only asymptotic
states, like in (\ref{sct}).  We denote Fermi wave numbers
$k_n=\sqrt{2m(E_F-E_n)}/\hbar$ and Fermi velocities $v_n=\hbar k_n/m$.
Only modes with $E_n<E_F$ contribute at zero temperature.

We construct an extended Hilbert space consisting of left and right
going states, $L$ and $R$, respectively. The relation between
operators in the reduced and standard space is
\ba
&&\hat{\psi}_{L\bar n}(x) = \int_{k<0}\frac{dkdx'dy}{2\pi}
e^{ik(x'-x)} \psi^\ast_n(y)\hat{\psi}_\sigma(x',y), 
\nonumber\\
&&\hat{\psi}_{R\bar{n}}(x) = \int_{k>0}\frac{dkdx'dy}{2\pi}
e^{ik(x'-x)} \psi^\ast_n(y)\hat{\psi}_\sigma(x',y). 
\ea
In fact, only $k\sim -k_n$ and $k\sim k_n$ play a role for $L$ and
$R$, respectively. The actual dynamics of states deep below or above
the Fermi sea can be ignored and its only reminder will be some
ultraviolet regularization or cutoff.  The new operators satisfy
anticommutation relations
\ba
&&\{\hat{\psi}_{A\bar n}(x),\hat{\psi}_{B\bar m}(x')\}=0,\nonumber\\
&&\{\hat{\psi}_{A\bar n}(x),\hat{\psi}^\dag_{B\bar m}(x')\}=\delta_{AB}\delta_{\bar n\bar m}\delta(x-x')
\ea
for $A,B=L,R$.

With help of the above stated approximations, we can write $\hat{H}$
in the new space as
\begin{eqnarray}
  &&\hat{H}_0+\sum_{\bar n} q_nv_n\left[\hat{\psi}^\dag_{L\bar
    n}(0)\hat{\psi}_{R\bar n}(0)+\hat{\psi}^\dag_{R\bar
    n}(0)\hat{\psi}_{L\bar n}(0)\right]+\label{ham2}\\
&&\sum_{\bar n}\int dx eV(x,t)\left[\hat{\psi}^\dag_{L\bar
    n}(x)\hat{\psi}_{L\bar n}(x)+\hat{\psi}^\dag_{R\bar
    n}(x)\hat{\psi}_{R\bar n}(x)\right],\nonumber 
\end{eqnarray}
where $\hat{H}_0$ is equal to
\begin{equation}
\sum_{\bar n} \int dxi\hbar v_n\left[\hat{\psi}^\dag_{L\bar
    n}(x)\partial_x\hat{\psi}_{L\bar 
    n}(x)-\hat{\psi}^\dag_{R\bar n}(x)\partial_x\hat{\psi}_{R\bar
    n}(x)\right].
\end{equation}
It is necessary to regularize the second term in Eq.~(\ref{ham2}) to
define the transmission and the reflection coefficients,
$T_n=\cos^2(q_n/\hbar )$ and $R_n=\sin^2(q_n/\hbar)$, respectively
(see the discussion in the appendix A).

The current operator (\ref{cur1}) in the new basis is replaced by
\begin{equation}
\hat{I}(x) = \sum_{\bar n}e
v_n(\hat{\psi}^\dag_{R\bar n}(x) \hat{\psi}_{R\bar n}(x)
-\hat{\psi}^\dag_{L\bar n}(x) \hat{\psi}_{L\bar n}(x)). 
\end{equation}
As we will later see, it is disadvantegous to operate directly with
the field operators. To circumvent this it is possible to introduce
bosonic operators (viz. quadratic forms of Fermion operators) and to
develop a closed algebra for those. Since current operators are
generally non-commuting, we have to cope with their algebra. To this
end, we introduce the following auxiliary operators
 \ba
&&\hat{I}_{0\bar n}(x_n)
=\frac{ev_n}{2}(\hat{\psi}^\dag_{L\bar n}(x) \hat{\psi}_{L\bar n}(x)
+\hat{\psi}^\dag_{R\bar n}(-x) \hat{\psi}_{R\bar n}(-x)),\nonumber\\
&&\hat{I}_{1\bar n}(x_n)
=\frac{ev_n}{2}(\hat{\psi}^\dag_{L\bar n}(x) \hat{\psi}_{L\bar n}(x)
-\hat{\psi}^\dag_{R\bar n}(-x) \hat{\psi}_{R\bar n}(-x)),\nonumber\\
&&\hat{I}_{2\bar n}(x_n)
=\frac{iev_n}{2}\hat{\psi}^\dag_{L\bar n}(x) \hat{\psi}_{R\bar n}(-x)
+\mathrm{h.c.}\;,\nonumber\\
&&\hat{I}_{3\bar n}(x_n)
=\frac{ev_n}{2}\hat{\psi}^\dag_{L\bar n}(x) \hat{\psi}_{R\bar n}(-x)
+\mathrm{h.c.}\;.\label{ijn}
\ea
with  $x_n=x/v_n$. Note that $x_n$ has the unit of time. It will be
more convenient for us than length units.
The Hamiltonian (\ref{ham2}) can be now written as
\ba
\hat{H} & = & \hat{H}_0+\sum_{\bar n}2q_n\hat{I}_{3\bar n}(0)/e\nonumber\\
&+&\sum_{\bar n,\pm}\int ds\;V(\pm sv_n,t)[\hat{I}_{0\bar
  n}(s)\pm\hat{I}_{1\bar n}(s)] \,.
\label{hscat}
\ea
The current takes the form
\begin{equation}
\hat{I}(x)=\sum_{\bar n}\left[\hat{I}_{0\bar n}(-x_n)-\hat{I}_{0\bar n}(x_n)-
\hat{I}_{1\bar n}(-x_n)-\hat{I}_{1\bar n}(x_n)\right].
\end{equation}
Using arguments similar to bosonization\cite{lut} (see appendix A), we
get the following commutation rule
\ba
&&[\hat{I}_{j\bar n}(s),\hat{I}_{k\bar m}(s')]=ie^2\delta_{\bar n\bar m}\delta_{jk}
\partial_s\delta(s-s')/4\pi\nonumber\\
&&
+\sum_li\epsilon_{jkl}e\delta_{\bar n\bar m}\hat{I}_{l\bar n}(s)\delta(s-s')\label{comm}
\ea
with $\epsilon_{jkl}$ equal $+1$ for $jkl=123,231,312$, $-1$ for
$jkl=321,213,132$ and zero otherwise. The last useful commutator is
\begin{equation}
[\hat{H}_0,\hat{I}_{j\bar n}(s)]=-i\hbar\partial_s\hat{I}_{j\bar n}(s)\label{comh}
\end{equation}
and one can in principle write
\begin{equation}
\hat{H}_0=2\pi\hbar\int ds\sum_{j\bar n}\hat{I}^2_{j\bar n}(s).
\end{equation}
The current $\hat{I}_0$ corresponds to the total charge changes in the
leads.  It is preserved due to absence of capacitive effects and
behaves like a free bosonic field.  Without the scattering term
$\hat{I}_3$ in (\ref{hscat}) only the first term in (\ref{comm}) needs
to be considered in the dynamics, since only $\hat{I}_0$ and
$\hat{I}_1$ are dynamical variables. The system could be fully
bosonized and all correlation functions can be calculated. A
nonvanishing scattering retains some fermionic features due to the
 commutator $[\hat{I}_1,\hat{I}_3]$. As a consequence the
fluctuations become non-Gaussian.

\subsection{Equilibrium averages}

Far from the junction and without bias potential, we can write the initial
density matrix as
\begin{equation}
\hat{\rho}=\exp(-\hat{H}_0/k_BT)/\mathrm{Tr}\exp(-\hat{H}_0/k_BT).
\end{equation}
The commutator (\ref{comh}) gives the useful relation
\begin{equation}
\label{eq:fluctdiss}
\mathrm{Tr}\hat{\rho}\hat{I}_{j\bar n}(s)\hat{A}=
\mathrm{Tr}\hat{\rho}\hat{A}\hat{I}_{j\bar n}(s+i\hbar/k_BT).
\end{equation}
A similar relation has been introduced in Ref. \onlinecite{tobis}.
For convenience we rewrite the previous equation and the relations
(\ref{comm}) in the frequency domain for the operators
\begin{equation}
\hat{I}_{j\bar n}(\omega)=\int d s\;e^{i\omega s}\hat{I}_{j\bar n}(t).
\end{equation}
From now on, a Greek argument will always denote the
Fourier-transformed operators.  The current algebra reads now
\ba
&&[\hat{I}_{j\bar n}(\alpha),\hat{I}_{k\bar m}(\beta)]=e^2\delta_{\bar
  n\bar m}\delta_{jk}
\alpha\delta(\alpha+\beta)/2\nonumber\\
&& +\sum_li\epsilon_{jkl}e\delta_{\bar n\bar m}\hat{I}_{l\bar
  n}(\alpha+\beta)\;.
\label{comm1}
\ea
Eq.~(\ref{eq:fluctdiss}) reproduces the fluctuation-dissipation theorem \cite{fdt}
\begin{equation}
\mathrm{Tr}\hat{\rho}\hat{I}_{j\bar n}(\omega)\hat{A}=e^{\frac{\hbar\omega}{k_BT}}
\mathrm{Tr}\hat{\rho}\hat{A}\hat{I}_{j\bar n}(\omega).\label{ther}
\end{equation}
With the above derived tools it is now straightforward to calculate
equilibrium averages. As examples we obtain
\ba
&&\mathrm{Tr}\hat{\rho}\hat{I}_{j\bar n}(\omega)=0,\label{avg}\\
&&\mathrm{Tr}\hat{\rho}\hat{I}_{j\bar n}(\alpha)\hat{I}_{k\bar m}(\beta)
=\frac{e^2\delta_{\bar n\bar m}\delta_{jk}}{4}\delta(\alpha+\beta)(w(\alpha)+\alpha),\nonumber\\
&&
\mathrm{Tr}\hat{\rho}\hat{I}_{j\bar n}(\alpha)\hat{I}_{k\bar m}(\beta)
\hat{I}_{l\bar p}(\gamma)=\frac{ie^3\epsilon_{jkl}\delta_{\bar n\bar m}\delta_{\bar n\bar p}}{8}\delta(\alpha+\beta+\gamma)\label{three}\nonumber\\
&&
\left[u(\beta)(w(\alpha)-w(\gamma))+w(\alpha)-w(\beta)+w(\gamma)+\alpha-\gamma\right],
\nonumber 
\ea
where we introduced
\begin{eqnarray}
u(\omega) & = & \coth(\hbar\omega/2k_BT)\,,\label{defuw}\\
w(\omega) & = & \omega u(\omega)\,,\nonumber\\
w(s) & = & -\lim_{\epsilon\to0}\mathrm{Re}\frac{\pi
  (k_BT/\hbar)^2}{\sinh^2(\pi sk_BT/\hbar+i\epsilon)}\,.\nonumber 
\end{eqnarray}
We note that $w(s=0)$ is singular and all practical calculations
need a proper regularization around $s=0$. This can for example be an
ultraviolet cutoff of the order of the Fermi energy in the
frequency domain. 

\subsection{Heisenberg equations}

To generalize the previous method to a nonequilibrium situation, it is
necessary to solve the Heisenberg equations for the various current
operators we have introduced. It is a major advantage of the current
algebra, we have introduced, that this can be done exactly.  From now
on, we switch to Heisenberg picture, so all operators will be
transformed as
\begin{equation}
i\hbar\partial_t\hat{A}(t)=[\hat A(t),\hat
H(t)]+i\hbar\tilde{\partial}_t \hat{A}(t).\label{heis} 
\end{equation}
Here the $\tilde\partial_t$ denotes explicit time dependence of the
operator.  Remarkable, the Heisenberg equations can be solved
completely for the operators $\hat{I}_{j\bar n}$. The details are
described in the appendix B and in the following we merely present the
results.  To distinguish between time and spatial coordinate (which
has the same unit, when divided by Fermi velocity) we will denote by
$\hat{A}(s,t)$ an operator at position $s$ and time $t$. We assume an
arbitrarily time-dependent bias potential $V(x,t)=\theta(-x)V(t)$ and
neglect some regularization issues around $x=0$ and $x\to-\infty$,
which can of course be properly handled using the procedures mentioned
previously. We find for the Heisenberg operators
\begin{eqnarray}
&&\hat{I}_{0\bar n}(s,t)=\hat{I}_{0\bar n}(t+s),\nonumber\\
&&\hat{I}_{1\bar n}(s,t)=\hat{I}_{1\bar n}(t+s)\mbox{ for }s>0,\\
&&\hat{I}_{1\bar n}(s,t)=(T_n-R_n)\hat{I}_{1\bar n}(t+s)\nonumber\\
&&-\frac{e^2}{2\pi\hbar}T_nV(t+s)
-2\sqrt{R_nT_n}\hat{J}^{\Phi}_{\bar n}(t+s)\mbox{ for }s<0\nonumber
\end{eqnarray}
with
\begin{equation}
\hat{J}^{\Phi}_{\bar n}(s)=\hat{I}_{2\bar n}(s)\cos\Phi(s)-
\hat{I}_{3\bar n}(s)\sin\Phi(s).\label{defj}
\end{equation}
and
\begin{equation}
\Phi(s)=-\int_{-\infty}^s dt\;eV(t)/\hbar.\label{defphi}
\end{equation}
Similar expressions can be obtained for $\hat{I}_{2\bar n}(s,t)$ and
$\hat{I}_{3\bar n}(s,t)$ and are presented in the appendix B. The big
advantage of the above equations is that the Heisenberg current
operators are represented by linear combinations of equilibrium
operators. Hence, their averages are straightforwardly obtained using
the results of the previous subsection.

We conclude this section by emphasizing the main results we have
obtained so far. The calculation of higher order cumulants of the current in
a mesoscopic junction using the standard scattering approach is quite
a demanding task. This is true in particular at finite frequency, when
operator ordering issues become even more prominent. The method, we
have developed in this section builds in an elegant way on a very
simple algebra of current operators (interestingly resembling the
usual angular momentum algebra). Our method allows to obtain in a
relatively straightforward manner arbitrarily ordered current
cumulants of (almost) any order.

\section{The measurement -- POVM}

We now turn to the main topic of the article, how the current
correlators at high frequencies can be measured and how the measurement
protocol itself influences the measurement. We will first treat the
measurement on a phenomenological basis using the so-called positive
operator-valued measure (POVM). In this formalism, the measurement is
described by Kraus operators, \cite{kraus} which we phenomenologically
assume to have the form
\begin{equation}
  \hat{K}[I] = \int D\phi \mathcal{T}e^{\int dt \left[i\phi(t)(\hat{I}_R(t)-I(t))/e
    -\phi^2(t)/\tau\right]}\;.\label{krau}
\end{equation}
Here $\mathcal T$ denotes time ordering and
\begin{equation}
\hat{I}_R(t)=\int dx\hat{I}(x,t)g(x-x_R)\label{irr}
\end{equation}
is current operator averaged over a spatial region near the point
$x_R>0$. By virtue of Naimark's theorem \cite{naim}, the completeness
relation $\int DI \hat{K}^\dagger[I]\hat{K}[I]=1$ is sufficient to
guarantee that the measurement corresponds to a usual projective
measurement in some extended Hilbert space including the
detector. 

Additionally, we assume as usual that the detector and the system are
initially uncorrelated.\cite{haenggi} The spatial resolution is taken
into account by a convolution function $g(x)=e^{-x^2/2\Delta
  x^2}/{\sqrt{2\pi}\Delta x}$, which is parametrized by the spatial
sensitivity $\Delta x$. To ensure, that the current on one side of the
junction is measured, we also take $\Delta x\ll x_R$. The Kraus
operator contains the parameter $\tau$, which plays the role of a
measurement sensitivity. Varying $\tau$, the measurement changes from a
weak, but nondemolishing measurement in the limit $\tau\to0$ to a strong
projective measurement for $\tau\to\infty$, which however yields not
the expected result due to a large disturbing noise of the
detector. For practical reasons, we convert distances into times,
namely, $\tau_n=\Delta x/v_n$ and $t_n=x_R/v_n$. The physical meaning
of $\tau_n$ is therefore the time, which an electron in the lead in
channel $n$ effectively interacts with the detector. On the other
hand, $t_n$ is related to the time-of-flight between the scatterer and
the detector.

The probability density of a given time trace for the current is given
by a POVM \cite{povm}
\begin{equation}
\rho[I]=\mathrm{Tr}\hat{\rho} \hat{K}^\dag[I]\hat{K}[I], 
\end{equation}
The generating functional is defined as
\begin{equation}
  \label{eq:cgf}
  {\mathcal S[\chi]} = \ln\left\langle \exp\left[i\int
      dt\chi(t)I(t)/e\right]\right\rangle_\rho\,
\end{equation}
where the average is defined as
\begin{equation}
  \label{eq:average}
  \left\langle \ldots \right\rangle_\rho=\int DI\rho[I]\ldots\,.
\end{equation}
Using the Kraus operators, one obtains for the generating functional
\begin{equation}
  \mathcal S[\chi]
  =\ln\int D\phi\: e^{S[\chi,\phi]-\int dt(2\phi^2(t)+\chi^2(t)/2)/\tau}\;,\label{genf}
\end{equation}
where $S[\chi,\phi]$ is the standard Keldysh functional, defined as
\begin{eqnarray}
S[\chi,\phi] & = & \ln\mathrm{Tr}\left\{\hat{\rho}
  \tilde{\mathcal T}\exp\left[\int
  \frac{idt}{2e}(\chi(t)+2\phi(t))\hat{I}_R(t)\right]\right.\nonumber\\ 
&&\times\left.
  \mathcal T\exp\left[\int \frac{idt}{2e}(\chi(t)-2\phi(t))
    \hat{I}_R(t)\right]\right\}.\label{genf1}
\end{eqnarray}  
The measure $D\phi$ is scaled to keep $\mathcal S[\chi\equiv 0]=0$.

To calculate averages we need the transformations presented in
Appendix C. 
For future convenience we shall denote $s_{\pm n}=s\pm t_n$ and
\begin{equation}
\theta_n(s)=\theta(s)h'_n(s),\:
h_n(s)=e^{-\frac{s^2}{4\tau_n^2}}/{2\sqrt{\pi}\tau_n}.
\label{tndef}
\end{equation}
The mean current is
\begin{equation}
\langle I(t)\rangle_\rho=-ie\left.\frac{\delta\mathcal
    S}{\delta\chi(t)}\right|_{\chi\equiv 0} 
=\frac{e^2}{\pi\hbar}\sum_{n}V(t_{-n})T_n\,.
\end{equation}
Hence, the conductance is robust against detector backaction as it
does not depend on $\tau$. 
In the limit of no-delay measurement ($t_n\to 0$), the measured
current follows the (time-dependent) voltage
\begin{equation}
\langle I(t)\rangle_\rho=GV(t)\;,\;G=\frac{2e^2}{h}\sum_{n}T_n.\label{cond}
\end{equation} 
The correlation function is
\begin{equation}
\langle\delta I(a)\delta I(b)\rangle_\rho=\left.\frac{-e^2\delta^2\mathcal S}{\delta\chi(a)\delta\chi(b)}\right|_{\chi\equiv 0}
=e^2P(a,b)
\end{equation}
for $\delta I=I-\langle I\rangle_\rho$.
We are interested in frequency scales of the current fluctuations
in the regime $\omega\ll\tau_n^{-1}$. Note, that the scale $1/\tau_n$
plays a role of the maximal bandwidth of the detector. This means that
the detector instantaneously measures the current, which is a
reasonable assumption also in typical experiments. 

In this case (see details in the appendix C) the noise measured at the
detector contains several contributions and can be written as
\begin{equation}
P(a,b)=\frac1\tau\delta(a-b)+P_\tau(a-b)+P_0(a-b)+P_e(a,b).\label{pnoise}
\end{equation}
The first term is just a Gaussian white noise of the detector, which
one would also expect classically.
The Fourier transforms of the second term  can be written as
\ba
&&P_\tau(\omega)=
\frac{\tau}{4\pi^2}\left|\sum_{n}(2i\theta_n(\omega)+\omega R_ne^{2i\omega t_n})\right|^2,\nonumber\\
&&\theta_n(\omega)\simeq -h_n(0)
-i\omega/2\label{ptau}
\ea
with $\theta_n$ defined by (\ref{tndef}). Therefore, $P_\tau$ has a
frequency dispersion solely determined by the properties of the
detector. It is independent of the bias voltage and can, thus, add an
unknown contribution to
the voltage independent background noise. Note, that it can be also
made negligible by a suitable choice of the detector parameters,

The third term has the form
\begin{equation}
P_0(\omega)=
\frac{w(\omega)}{\pi}\sum_{n}
(T_n^2+R_n(1-\cos(2\omega t_n)).\label{poo}
\end{equation}
Here $w(\omega)$ is defined by Eq.~(\ref{defuw}). This term is also
independent of the bias voltage and can be neglected at sufficiently
low temperatures. .

Finally the most interesting contribution to the noise is the voltage-dependent part
$P_e$, which we can write with the help of (\ref{phiav}) and
$P_e(a,b)=\sum_n P_n(a,b)$ as
\begin{equation}
P_n(a,b)=R_nT_n\frac{w(a-b)}{\pi\Gamma_n(a-b)}
\cos\varphi(a_{-n},b_{-n})
\label{exc}
\end{equation}
with $\Gamma_n(t)=\exp\left(\tau(h_n(0)-h_n(t))/4\right)$
and
\begin{equation}
\varphi(a,b)=\Phi(a)-\Phi(b)=\int_a^b dt\;eV(t)/\hbar.\label{defvphi}
\end{equation}
Still, this result can be applied to arbitrary frequency after multiplying
$P_n(\omega,\omega')$ by the damping factor $e^{-(\omega^2+\omega'^2)\tau_n^2/2}$.
The noise (\ref{exc}) contains the usual symmetrized quantum
noise in agreement with existing results.\cite{blanter,lesovik} The
factor $\Gamma$ results from the detector backaction on the current
during the measurement on a time-scale set by the phenomenological
parameter $\tau$.  
Note that $\Gamma(t\to 0)\to 1$. A high voltage $V$
leads to a strongly  
oscillating term $\cos(\varphi)$,  so that $P_n$ effectively probes
the short time-scale $t\sim\hbar/eV$. Hence, 
in the limit of high voltage we can take $\Gamma(\hbar/eV)\to 1$ and
the noise is independent of the detector's backaction. 
To illustrate this effect more clearly let us take $V =$ const and
$t_n=0$. In this case 
\ba
P_n(\omega,\omega') & = & 2\pi\delta(\omega+\omega')P_n(\omega),\nonumber\\
P_n(t) & = & R_nT_n\frac{w(t)}{\pi\Gamma_n(t)}
\cos(eVt/\hbar).
\ea
In Fourier space we obtain the convolution
\begin{eqnarray}
  P_n(\omega) & = & \int \frac{d\alpha}{(2\pi)^2}
  R_nT_n\Gamma_n^{-1}(\omega-\alpha)\\ \nonumber
  &&\times\left[w(\alpha+eV/\hbar)+w(\alpha-eV/\hbar)\right].
\label{dcnoise}
\end{eqnarray}
Due to the symmetry of $\Gamma(\omega)=\Gamma(-\omega)$ and the fact that
$\Gamma_n^{-1}(\omega\to\infty)$ vanishes, we can replace
$\Gamma^{-1}_n(\omega)\to 2\pi\delta(\omega)$ at high voltage and
get the shot noise $P_n=R_nT_n|eV|/\pi\hbar$. Let us find the
backaction corrections to the shot noise at zero frequency
($\omega=0$),
\begin{equation}
P_n(0)=\int \frac{d\alpha}{2\pi^2} R_nT_nw(\alpha)\Gamma_n^{-1}(\alpha+eV/\hbar).
\end{equation}
For $\tau\ll\tau_n$, have
\begin{equation}
\Gamma_n^{-1}(\alpha)=(2\pi-\sqrt{\pi}\tau/4\tau_n)\delta(\alpha)+
\tau e^{-\alpha^2\tau_n^2}/4
\end{equation}
and finally at zero temperature ($w(\alpha)=|\alpha|$) we get
\begin{equation}
P_n(0)=R_nT_n[|eV|/\pi\hbar
+q(|eV|\tau_n/\hbar)\tau/8\pi^2\tau_n^2]
\end{equation}
with $q(x)=e^{-x^2}-2x\int_x^\infty dz\:e^{-z^2}$. 

\begin{figure}
  \includegraphics[scale=.65,clip=true,angle=270]{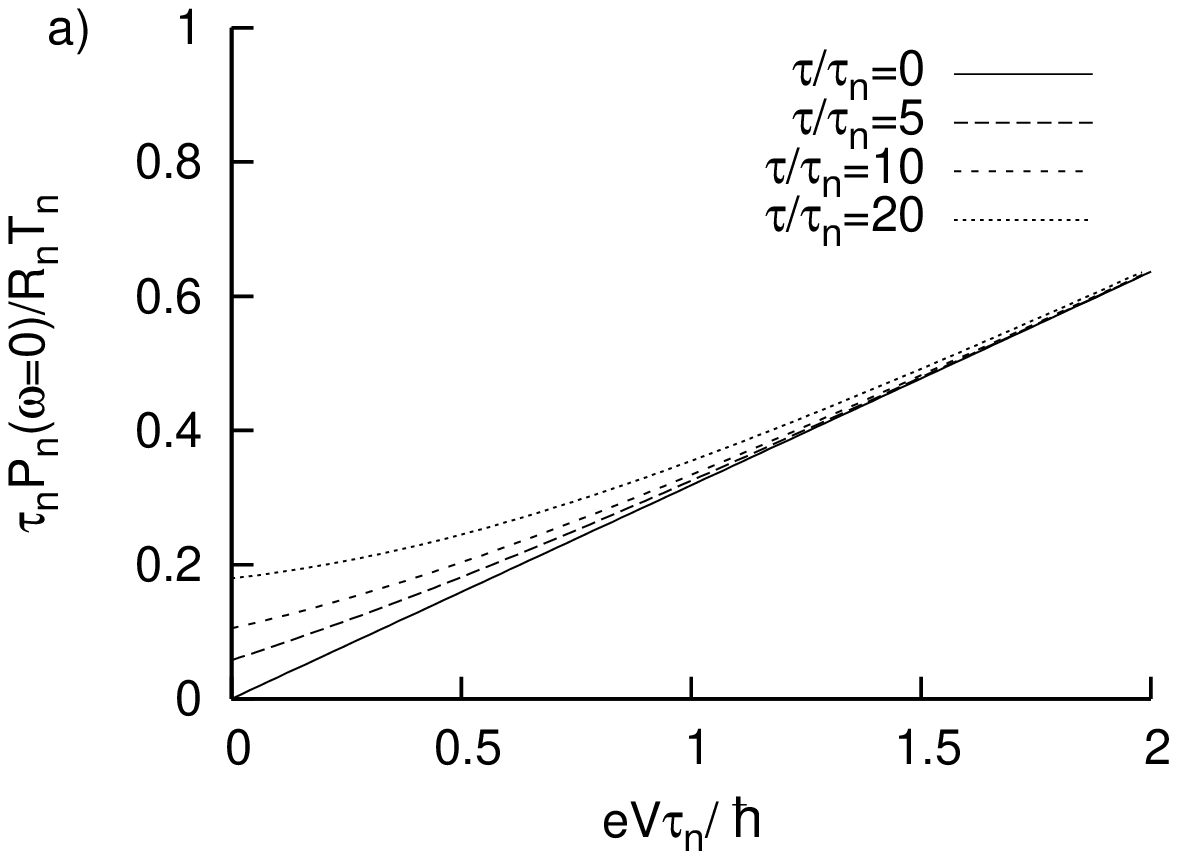}
  \includegraphics[scale=.65,clip=true,angle=270]{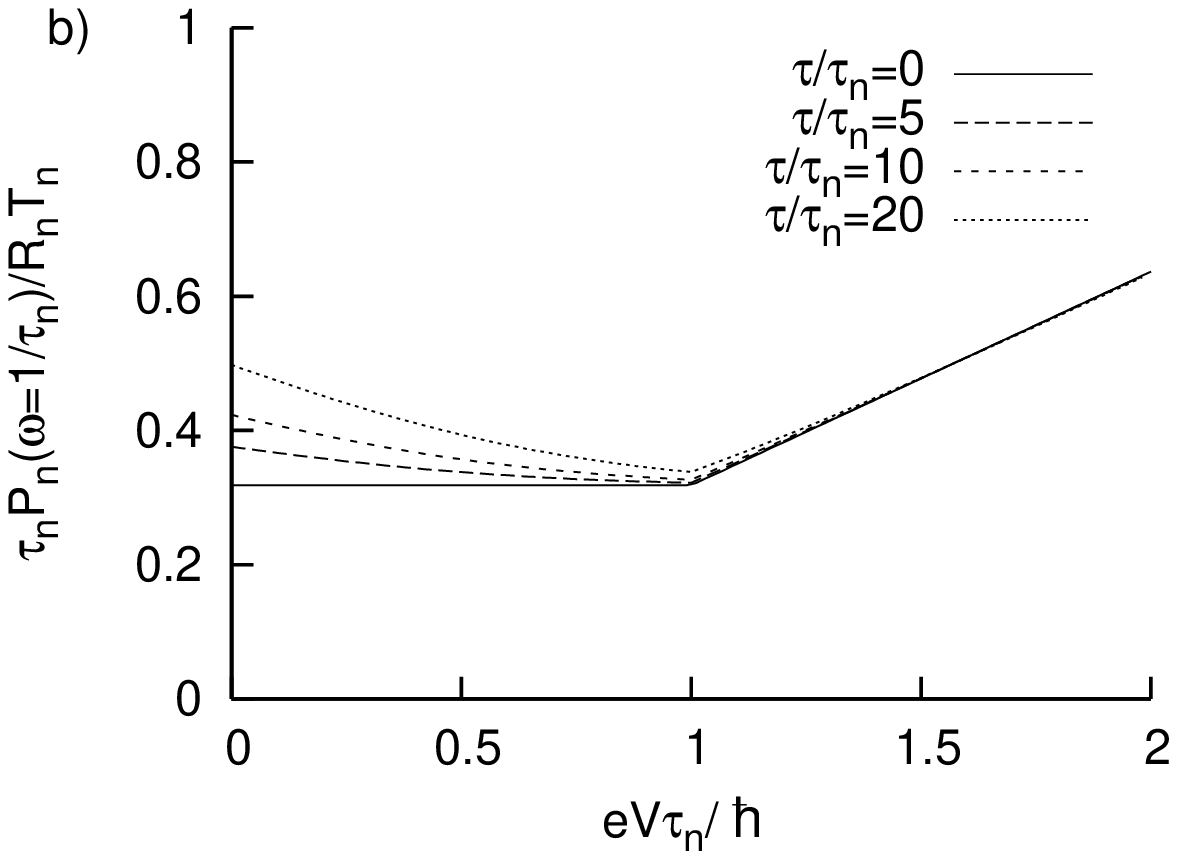}
  \caption{The voltage-dependent noise $P_n$ at dc voltage and zero temperature
    for zero frequency $\omega=0$ (a) and a finite frequency
    $\omega=1/\tau_n$ (b).  In both cases the additional noise due to
    the backaction of the detector leads to drastic changes in the
    voltage dependence for $eV\lesssim 1/\tau_n$. The magnitude of the
    noise increase depends on the ratio $\tau/\tau_n$. For larger voltages
    the usual shot noise is recovered.
  }\label{pof}
\end{figure}

To illustrate the effect of the backaction due to the detection
process we show in Fig.~\ref{pof} the low- and the high-frequency noise for different
detector parameters $\tau$. The upper panel shows that the effect of the
detector is strongest for small voltages $eV\lesssim
\hbar/\tau_n$. We can interpret this as follows: The additional noise added by the detection is dominant
as long as the electrons flow with a rate smaller than the inverse interaction
time with the detector $\lesssim 1/\tau_n$ through
the contact.
The influence of the backaction becomes negligible if the
electrons flow at a higher rate, so that they do not feel the
backaction and the full shot noise is recovered. A similar picture
holds if the detector is sensitive to the finite-frequency current
correlations. The signature of the quantum transport at
$\hbar\omega=eV$ is gradually smeared out if the detector becomes
slower, viz. $\tau\gtrsim\tau_n$. Interestingly, the noise first
decreases with a finite voltage. Finally we note, that Fig.~\ref{pof}
shows numerically that the correction is small even at
$\tau\sim\tau_n$.

\subsection{Many detectors}
We can generalize the Kraus operator (\ref{krau}) to the case of many detectors
\begin{equation}
  \hat{K}[I] = \int D\phi \mathcal{T}e^{\int dt \sum_A(i\phi_A(t)(\hat{I}_A(t)-I_A(t))/e
    -\phi_A^2(t)/\tau_A)}\;.\label{krau2}
\end{equation}
In particular, we take two independent detectors placed on the left and right hand side of the QPC.
They are described similarly as in the single detector case, namely
\begin{equation}
\hat{I}_A(t)=\int dx\hat{I}(x,t)g_A(x-x_A),\mbox{ for }A=L,R\,.
\end{equation}
Here we assumed $x_R>0$ and $x_L<0$ and $g_A(x)=e^{-x^2/2\Delta x_A^2}/{\sqrt{2\pi}\Delta x_A}$ with $\Delta x_A\ll |x_A|$. The
latter assumption means that the measuring regions of the two
detectors do not overlap.  We also denote the interaction time of the
detector with the electrons as $\tau_{An}=\Delta x_A/v_n$.

The conductance (\ref{cond}) remains unchanged but the noise can be
now measured in four different ways.  Namely, we can define the four correlators 
\begin{equation}
  \langle\delta I_A(a)\delta I_B(b)\rangle_\rho = 
  -\left.\frac{e^2\delta\mathcal S}{\delta\chi_A(a)\delta\chi_B(b)}\right|_{\chi\equiv 0}
  =e^2P_{AB}(a,b)
\end{equation}
with $A,B=R,L$.

We shall consider only frequency scales $\omega\ll v_n/|x_A|$, so that
we can use $t_n=0$ in all derivations. Then, similarly as in the previous
section (details in the appendix C), we obtain expressions for all
correlators of the form
\begin{eqnarray}
  P_{AB}(a,b) & = & \frac{1}{\tau_A}\delta_{AB}\delta(a-b)
  +P_{AB}(a-b)\nonumber\\ &&+P_{0}(a-b)+P_e(a,b)\,.
\end{eqnarray}
Here, $P_0$ is defined by Eq.~(\ref{poo}) with $t_n=0$.
The excess noise $P_e$ is defined by Eq.~(\ref{exc}), but with
\begin{equation}
\Gamma_n(t)=\exp\left(\sum_A\tau_A(h_{An}(0)-h_{An}(t))/4\right)\,.
\end{equation}
Note that the excess contribution is the same for all correlators and
depends on the parameters of both detectors.
The voltage independent contributions differ for the auto-correlators
and the cross-correlators. We have
\begin{equation}
P_{AA}(\omega)=P_{\tau A}(\omega)+\frac{\omega^2\tau_B}{4\pi^2}\left(\sum_{n}T_n\right)^2\,,
\end{equation}
where $P_{\tau A}$ is defined by (\ref{ptau}) with $\tau=\tau_A$ and $\tau_n=\tau_{An}$
and $B=L,R$ for $A=R,L$, respectively. Finally, for $A\neq B$ we find
\begin{eqnarray}
&&P_{AB}(\omega)=\frac{\omega}{4\pi^2}\sum_{n}T_n
\sum_{m}\label{mlast}\\
&&\left[\tau_A(2i\theta_{Am}(\omega)-R_m\omega)+\tau_B(-2i\theta^\ast_{Bm}(\omega)-
R_m\omega)\right]\,.\nonumber
\end{eqnarray}

The possibility to measure several independent correlators has
interesting consequences.  In the case of a single detector one could in
practice measure only the \emph{voltage-dependent} contribution of the
noise as the offset noise has generally an unknown value and is
subtracted.  However, the use of two independent detectors helps to
estimate the background noise in the auto-correlation signal.
Hence, comparing cross- and auto-correlations we also can get a rough
estimate of the offset noise. Furthermore, in the limit $\tau_{L,R}\to
0$ we get $P_{LR}=P_0+P_e$. As $P_e$ is independently known from the
auto-correlation measurement, we can get information about the
voltage-independent part of the noise $P_0$.


\subsection{Higher cumulants}

Now we consider the effect of our detection scheme on the third and
fourth cumulants of the current fluctuations. While these are harder
to measure than noise correlations, they contain a tremendous deeper
information that the current. It should be noted, that the third
cumulant already at zero frequency contains a non-trivial ordering of
the current operators \cite{fcs} and we can expect a similar
non-trivial effect of the detection scheme. Quite generally the third
and forth cumulants are defined as
\begin{eqnarray}
  \langle\langle ABC\rangle\rangle & = & 
  \langle\delta A\delta B\delta C\rangle,\\
  \langle\langle ABCD\rangle\rangle & = & \langle\delta A\delta B\delta C\delta D\rangle
  -\langle\delta A\delta B\rangle\langle\delta C\delta D\rangle\nonumber\\
  &&\nonumber
-\langle\delta A\delta C\rangle\langle\delta B\delta D\rangle
-\langle\delta A\delta D\rangle\langle\delta C\delta B\rangle\,.
\end{eqnarray}
Here $\delta X=X-\langle X\rangle$ are fluctuations of some
observable. In the following these will be current fluctuations
operators at different times. 

We now apply out model of a detector, which is parametrized by a Kraus
operator to calculate the cumulants. Unfortunately, the general
expressions are too cumbersome to be shown here and we discuss only a
limiting case below.  Hence, we take the limit $|\tau_n\omega|,\tau/\tau_n\ll
1$, where $\omega$ describes relevant frequency scale of the measurement. In this case the only effect of the detector is a large
white Gaussian offset noise $1/\tau$. It only adds as a constant to 
the second cumulant, while higher cumulants are unaffected.

The derivation of the third cumulant is given in the appendix D. The result for
$C_3(a,b,c)=\langle\langle I(a)I(b)I(c)\rangle\rangle$ is
\begin{eqnarray}
  &&C_3(a,b,c)=
e^3\sum_{n}
\left\{\vphantom{\sum_{\sigma(abc)}}R_nT_n(R_n-T_n)\times\right.\nonumber\\
&&\frac{eV(a_{-n})}{\pi\hbar}\delta(a-b)\delta(a-c)+\sum_{\sigma(abc)}\int ds\:P_n(a,b;s)\times\nonumber\\
&&\left[(2T_n-1)w(s-c_{-n})+w(s-c_{+n})\right]\pi/2
\left.\vphantom{\sum_{\sigma(abc)}}
\right\}\label{third2}
\end{eqnarray}
where the summation is over all permutations of the set $abc$. 
At zero flight time and either zero temperature and frequency
or tunneling limit only the first term survives.
The last part contains the so-called noise susceptibility \cite{gabel}
\begin{equation}
P_n(a,b;s)=\frac{\hbar \delta P_n(a,b)[V]}{e\delta V(s)}\label{susdef}\,,
\end{equation} 
generalized here to an arbitrary time-dependence of the bias.
The result (\ref{third2}) in the limit of zero flight time has a somewhat simpler
form than the existing results.
\cite{zaikin1}  The identification of the
noise susceptibility in the third cumulant has important consequences
on its detection. A general detector, which is equivalent to an electromagnetic environment,
and weakly coupled here, gives corrections to (\ref{third2}) of the form
\cite{reulet, reznikov, regab, been}
\begin{eqnarray}
&&C^{en}_3(\omega_1,\omega_2,\omega_3)=\sum_{\sigma(ijk)} \int d
\omega P(\omega_i,\omega_j;\omega)\times\label{encor}\\ 
&&[b_0(\omega)w(\omega)\delta(\omega-\omega_k)+
b_1(\omega) P(-\omega,\omega_k)/e^2]/4.\nonumber
\end{eqnarray}
Here $P(\alpha,\beta;s)=\delta P(\alpha,\beta)/\delta V(s)$ is the
noise susceptibility. The function
$b_0$ represents the influence of environmental noise and $b_1$ is the
voltage-dependent feedback of the environment due to the noise of the
system. They can in principle be modeled by an
effective electric circuit. In practice, the precise environmental
circuit is not known, and the functions are determined by fitting some
model environment.\cite{regab}

\begin{figure}[ht]
\includegraphics[scale=.55,clip,angle=270]{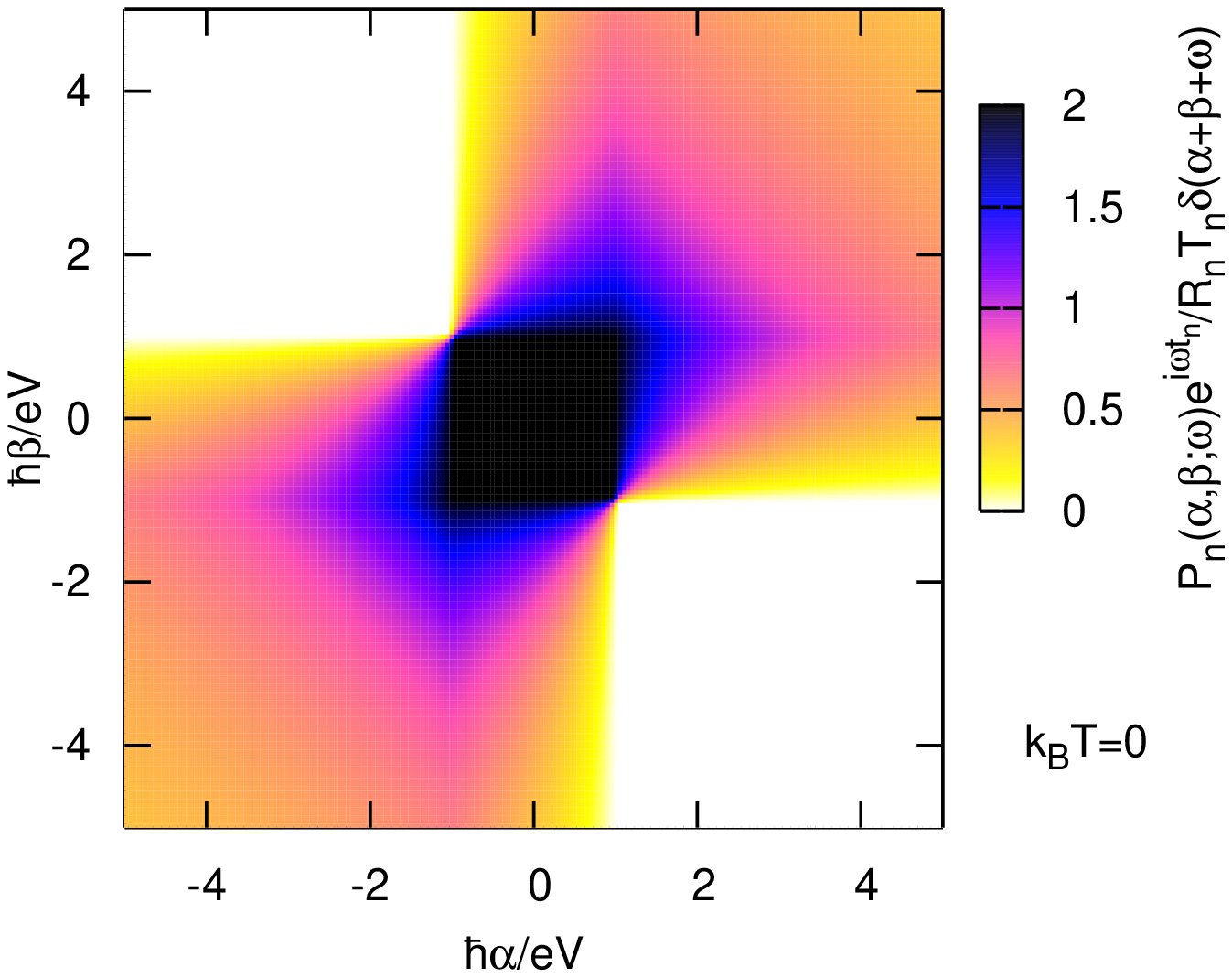}
\includegraphics[scale=.55,clip,angle=270]{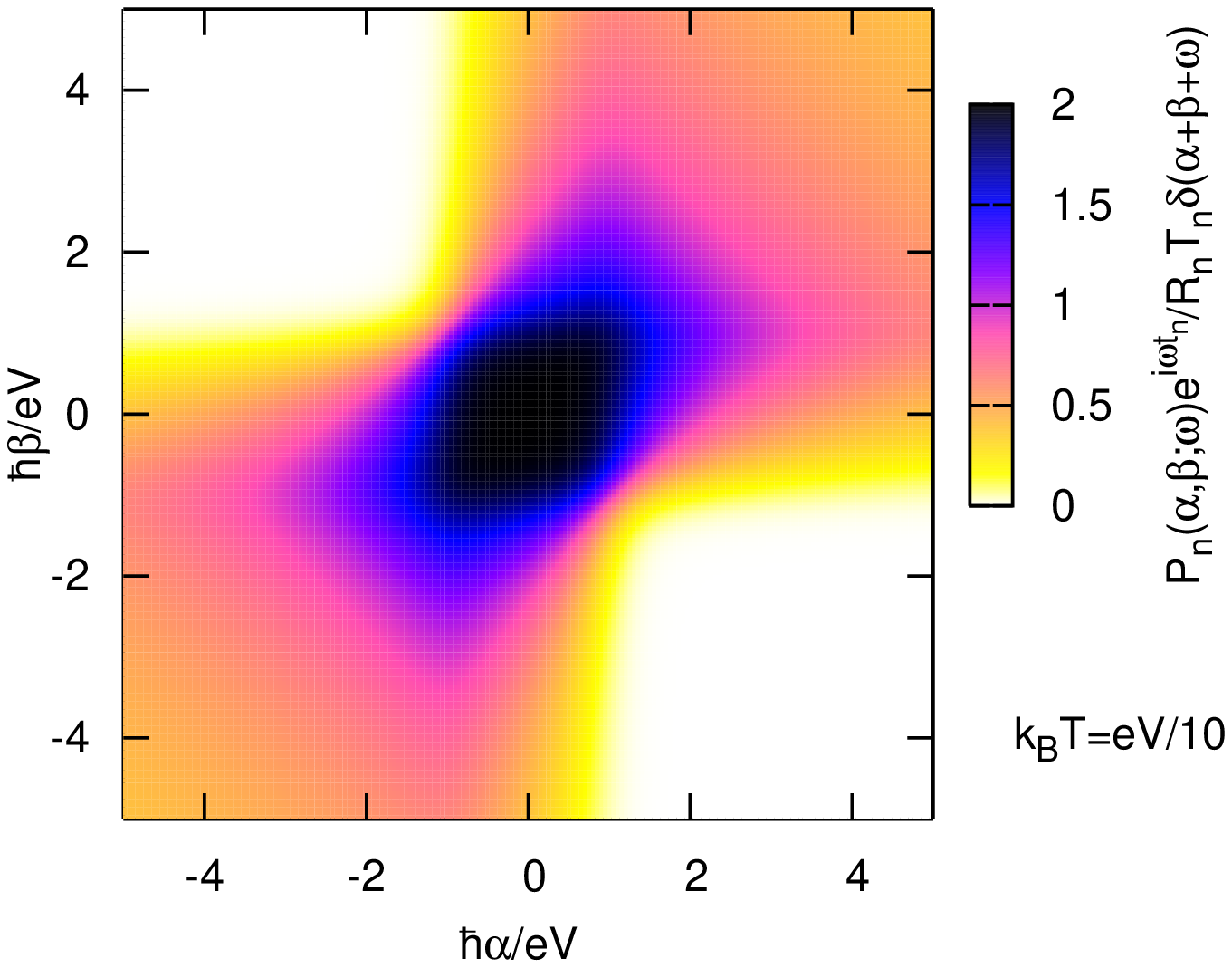}
\includegraphics[scale=.55,clip,angle=270]{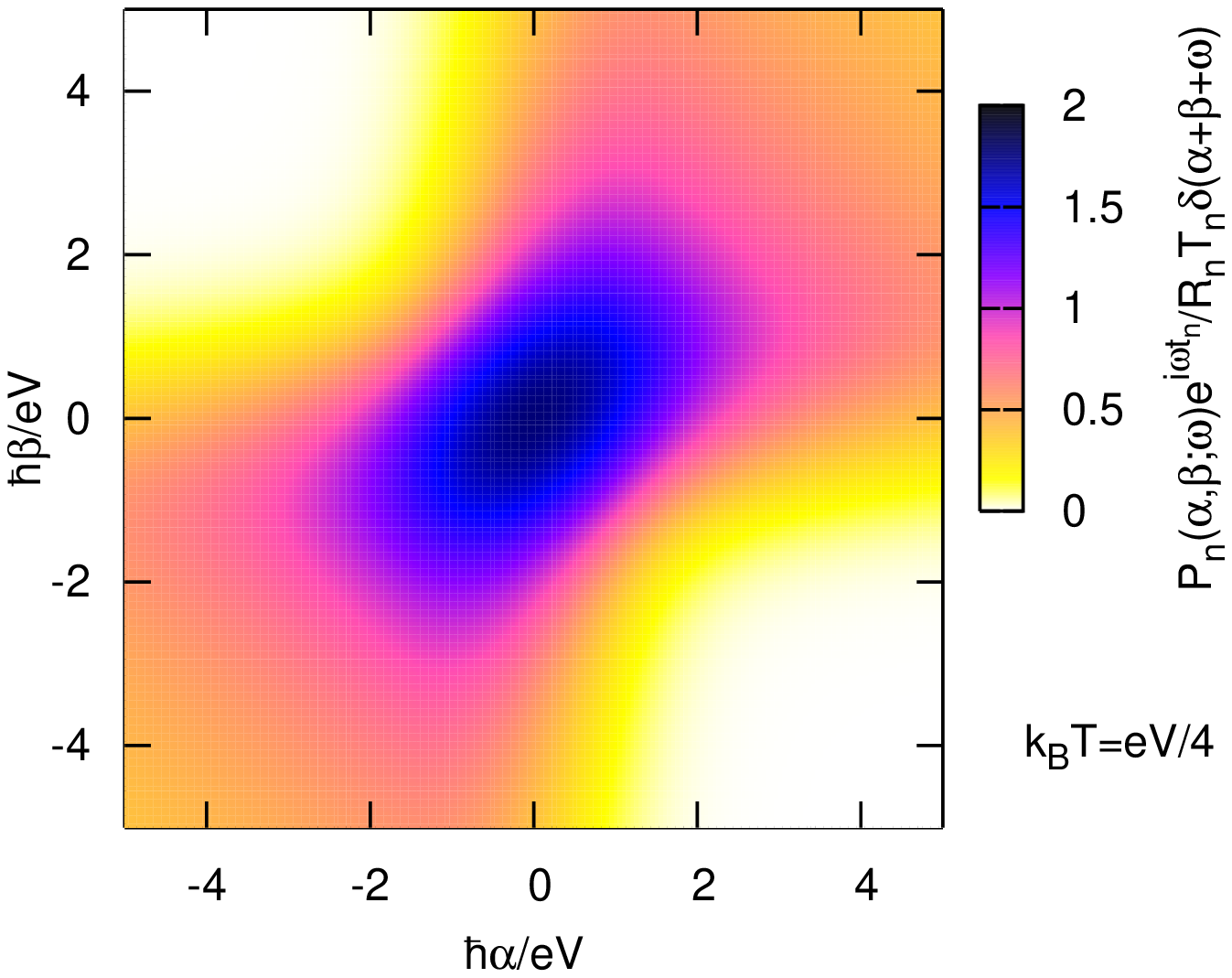}
\caption{(Color online) The susceptibility $P_n(\alpha,\beta;\omega)$ at dc voltage
  defined in (\ref{pndc}) for different temperatures.  The cross
  symmetry is preserved for all temperatures.}\label{pns}
\end{figure}

An important case in practice is a constant bias voltage, $V=$
const. The expressions for the noise susceptibility and the third
cumulant are simplified considerably. The noise susceptibility turns
out to be
\begin{eqnarray}
  P_n(\alpha,\beta;\omega) &=
  &R_nT_n\delta(\alpha+\beta+\omega)e^{-i\omega t_n}\times
  \label{pndc}
  \\
  &&\sum_{\gamma=\alpha,\beta} [w(\gamma - eV/\hbar)-w(\gamma +
  eV/\hbar)]/\omega\,.
  \nonumber
\end{eqnarray}
The noise susceptibility is shown in Fig.~\ref{pns} for different
temperatures. In all cases, it preserves the symmetry
$\alpha\leftrightarrow\pm\beta$. At zero temperature, a non-analyticity occurs along lines $|eV/\hbar|=|\alpha|,|\beta|$,
which is smeared out at increasing temperatures. 

\begin{figure}[ht]
\includegraphics[scale=.55,clip,angle=270]{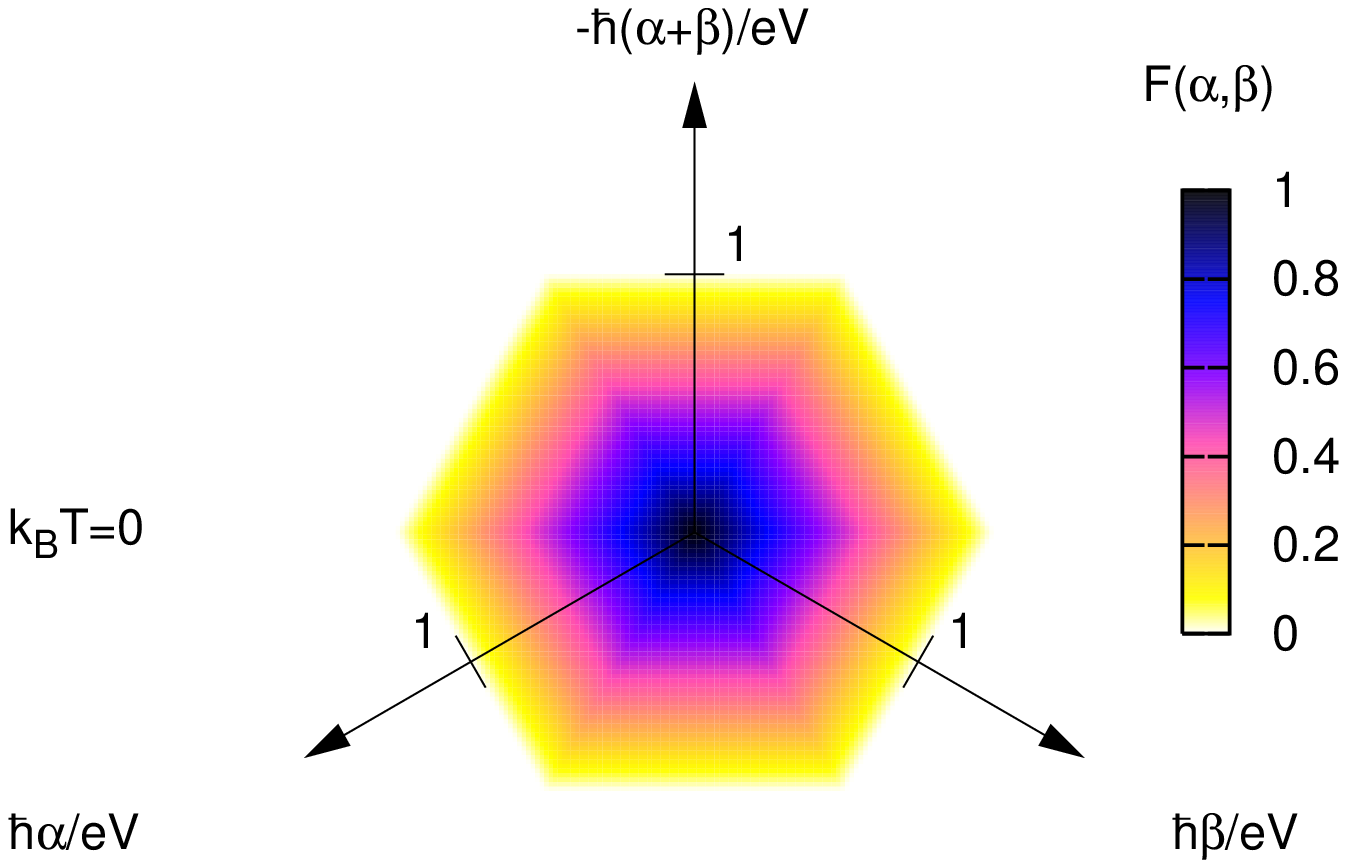}
\includegraphics[scale=.55,clip,angle=270]{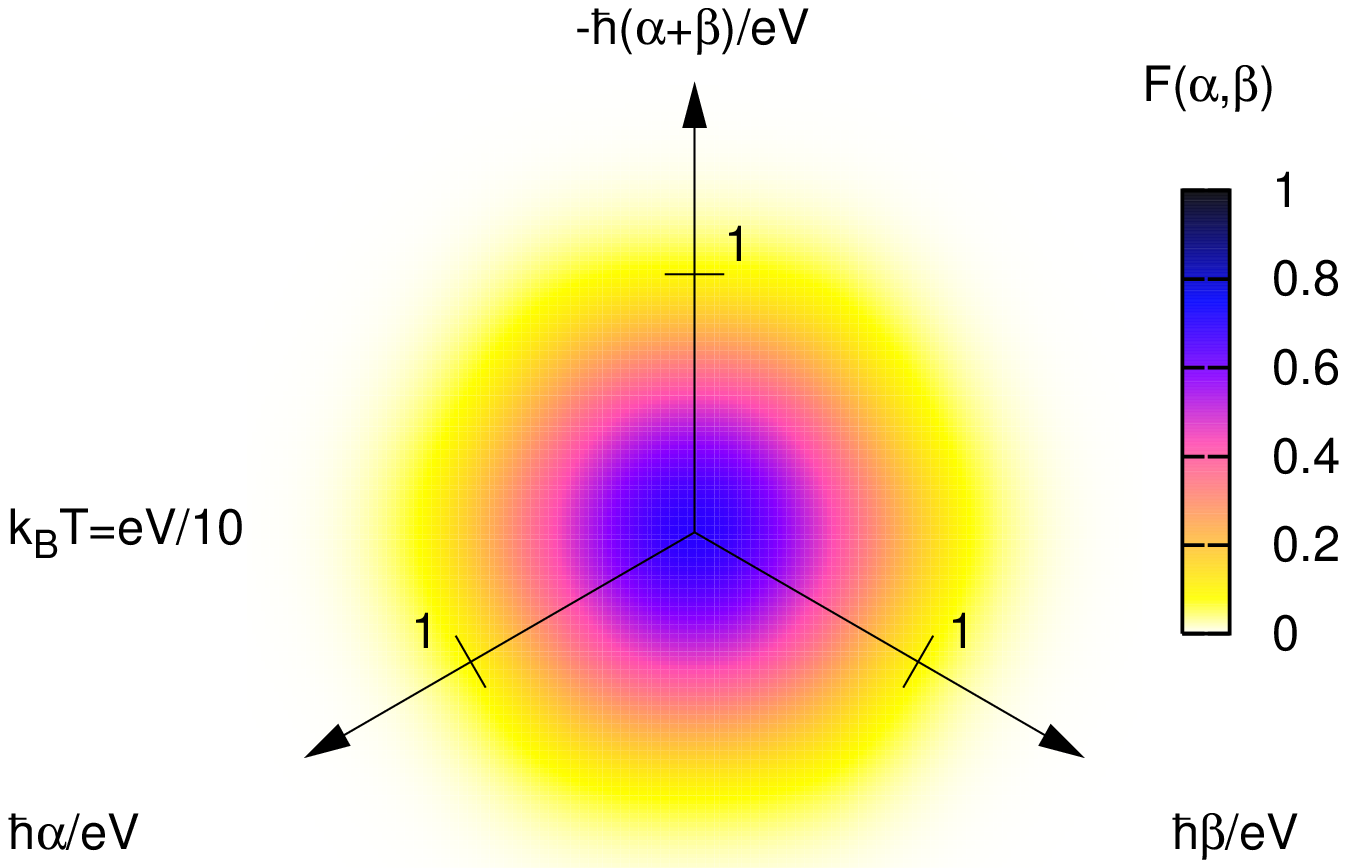}
\includegraphics[scale=.55,clip,angle=270]{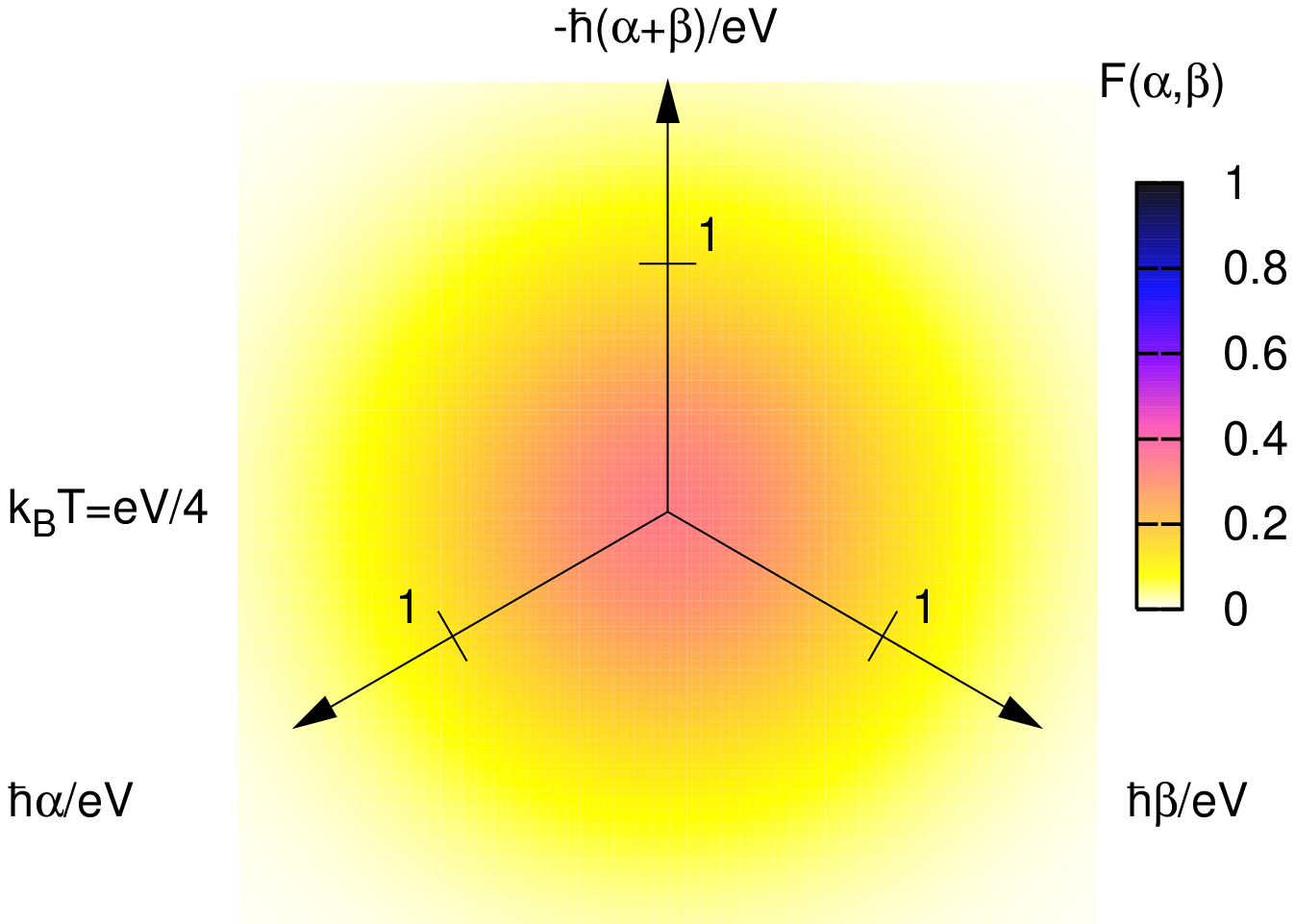}
\caption{(Color online) The factor $F(\alpha,\beta)$ defined in (\ref{sb3}) for
  different temperatures.  The hexagonal symmetry is preserved for all
  temperatures.}\label{qqfig}
\end{figure}

The third cumulant also becomes much simpler. Due to the relation
$C_3(\alpha,\beta,\omega)=2\pi\delta(\alpha+\beta+\omega)\bar{C}_3(\alpha,\beta)$,
it effectively  depends only on the frequencies $\alpha$ and
$\beta$. The delta function imposes the constraint $\alpha+\beta+\omega=0$ so only two of the three frequencies
are independent. The cumulant is hence symmetric under changes $\alpha,\beta\leftrightarrow\omega=-\alpha-\beta$.
In the following expressions, we either use two independent arguments ($\alpha$,$\beta$) or three
constrained
($\alpha$,$\beta$,$\omega$).
For a negligible flight time $|\alpha|,|\beta|,|\omega|\ll t_n^{-1}$ we have
\begin{eqnarray}
  \bar{C}_3(\alpha,\beta) & = & \sum_n R_nT_n(1-2T_nF(\alpha,\beta))2e^4V/h,\nonumber\\
  F(\alpha,\beta) & = & 1-\sum_{\sigma(\alpha\beta\omega)}
  u(\omega)\hbar/eV\times\label{sb3}\\
  &&[w(\alpha - eV/\hbar)-w(\beta + eV/\hbar)]/4\,,\nonumber
\end{eqnarray}
where we sum over permutations of the constrained set $\alpha\beta\omega$ .
The temperature-dependence is encoded in a single universal function $F$,
which does not depend on the channel transparency.
In agreement with previous results \cite{zaikin1,salo} we find
at zero temperature
\begin{equation}
  F(\alpha,\beta) = 1-\mathrm{min}\{\mathrm{max}\{|\alpha|,
  |\beta|,|\alpha+\beta|\},|eV/\hbar|\}\hbar/|eV|.\label{qzero}
\end{equation}
The zero-frequency limit is of course \cite{levrez}
\begin{equation}
  F(0,0)=1-3\frac{\sinh U-U}{U(\cosh U-1)},\:U=eV/k_BT\,.\label{qzero1}
\end{equation}
The frequency dependence of the function $F$ is plotted in
Fig. \ref{qqfig}. The form of the plot is motivated by the 
symmetry of arguments and the constraint
$\alpha+\beta+\omega=0$. However, experimentally a verification of the full
frequency dependence is a desirable challenge.

The effect of flight times on the second cumulant in Eq.~(\ref{poo})
was only an additional offset noise,
independent of bias voltage. In the case of the third cumulant the
terms, which depend on the flight time, depend on voltage, see Eq.~(\ref{third2}).
For $\omega,\omega'\gg t_n^{-1}$ we get
\begin{equation}
\bar{C}_3(\omega,\omega')=F(\omega,\omega')\bar{C}_3(0,0)
\end{equation}
so the cumulant drops to zero at large frequencies, contrary to the zero flight time case, where it
stays nonzero and proportional to $eV$. 
The recent experiment \cite{regab} does not show any reduction of the
third cumulant at high frequency, so the frequency scale defined
by the flight time must still be beyond experimental reach.  However,
there is an environmental correction of a similar magnitude. We will continue the discussion of
the third cumulant in Sec. IV.

Lastly, the fourth cumulant can be evaluated in the similar way.
Here we present only the result at $eV=0$,
\begin{eqnarray}
&&\langle\langle
I(\omega_1)I(\omega_2)I(\omega_3)I(\omega_4)\rangle\rangle=\\
&&\sum_{n}\frac{e^4R_nT_n}{8}\delta\left(\sum_i\omega_i\right)\sum_{\sigma(\omega_i)}e^{2it_n(\omega_1+\omega_2)}u(\omega_3)u(\omega_4)\times\nonumber\\
&&[w(\omega_1)+w(\omega_2)-w(\omega_1+\omega_3)
-w(\omega_2+\omega_3)]\nonumber
\end{eqnarray}
where we perform summation over all permutations of the set $\omega_1,\dots,\omega_4$.
Due to hyperbolic identities, for $t_n=0$ it reduces to 
\begin{equation}
\sum_{n}\frac{e^4R_nT_n}{2}\delta\left(\sum_i\omega_i\right)\sum_i w(\omega_i).
\end{equation}
which agrees with existing results at small reflection or transmission. \cite{zaikin1}
One can in principle use this method to find analytic expressions for cumulants of any order, 
which is beyond the scope of this work. 

\subsection{Tunneling and transparent limit}

In some limits the generating function $\mathcal S[\chi]$ can be evaluated exactly.
For $\tau\to 0$, we can make the separation
\begin{equation}
\mathcal S[\chi]=S[\chi,0]-\int dt\;\chi^2(t)/2\tau,\nonumber
\end{equation}
where $S[\chi,\phi]$ is defined by (\ref{gene}). 
We are particularly interested in the tunneling limit $T_n\ll 1$ and
the transparent limit
$R_n\ll 1$ up to terms of the first order in $T_n$ ($R_n$).\cite{zaikin1} 
With help of appendix E, we get for $R_n\ll 1$ 
\begin{equation}
S[\chi,0]=2NS_0[\chi]+\int \frac{ieV(t)}{\pi\hbar}\sum_n\chi(t_{+n})+\sum_nS_{Rn}[\chi],
\end{equation}
and for $T_n\ll 1$ (tunneling limit),
\begin{equation}
S[\chi,0]=\sum_n 4S_0[\chi_{Tn}]+\sum_n S_{Tn}[\chi]\,.
\end{equation}
Here, $N$ denotes the number of modes,
\begin{equation}
S_{0}[\xi]=-\int dtdt'
\frac{w(t-t')}{4\pi}\xi(t)\xi(t'),\label{s00}
\end{equation}
is the generating functional of Gaussian vacuum fluctuations and
\begin{eqnarray}
&&S_{An}[\chi]=-\int \epsilon_A\frac{ieV(t)}{\pi\hbar}A_n\sin\chi(t_{+n})dt+\nonumber\\
&&\int dtdt'\frac{w(t-t')}{\pi}A_n\left\{\vphantom{\frac{1}{2}}\epsilon_A\chi_{An}(t)
\sin\chi(t'_{+n})\right.\label{san}\\
&&\left.-2\sin\frac{\chi(t_{+n})}{2}\sin
\frac{\chi(t'_{+n})}{2}e^{i\varphi(t',t)+D(t',t)[\chi_{An}]}\right\}\nonumber\,.
\end{eqnarray}
The arguments depend on the transmission (reflection) probabilities
and are defined by $2\chi_{An}(t)=\chi(t_{-n})+\epsilon_A\chi(t_{+n})$, $\epsilon_{R/T}=+/-$.
The kernel in the exponent is defined as
\begin{eqnarray}
&&D(a,b)[\xi]=\int_b^ads\int dt\;w(s-t)\xi(t)\label{ddd}\\
&&=\int (e^{i\omega b}-e^{i\omega a})iu(\omega)\xi(-\omega)\frac{d\omega}{2\pi}=\frac{k_BT}{\hbar}\int dt\:\xi(t)\times\nonumber
\\
&&\mathrm{Re}\left[\coth\left(\frac{\pi(a-t)}{\hbar/k_BT}+i\epsilon\right)-
\coth\left(\frac{\pi(b-t)}{\hbar/k_BT}+i\epsilon\right)\right].\nonumber
\end{eqnarray}

One can see that the most significant effect of the flight times in
the tunneling limit is Gaussian noise $S_0$ that is growing with distance
from the contact. This is because $\chi_T(t)=\chi(t_-)-\chi(t_+)$ vanishes at zero flight time
but for large flight times the two parts will become independent and not
cancel each other. The noise saturates to the same equilibrium value as in
the transparent limit. We stress again, however, that there is no
experimental evidence of reaching that timescale so far. The non-Gaussian part $S_{An}$ remains small
as it is proportional to $A_n$.

In the limit of a vanishing flight time $t_n=0$ the above formula
for the tunneling case simplifies to
\begin{eqnarray}
  \label{eq:quasi1}
  \frac{S[\chi,0]}{\sum_{n} T_n} && = 
  \int idt\;\sin\chi(t)\;eV(t)/\pi\hbar\label{gent}\\
  &&\!\!\!\!\!\!\!-\int dtdt'
\frac{2w(t-t')}{\pi}\sin\frac{\chi(t)}{2}\sin\frac{\chi(t')}{2}
\cos\varphi(t,t')\nonumber
\end{eqnarray}
with $w$ and $\varphi$ defined by (\ref{defuw}) and (\ref{defvphi}),
respectively.

It is tempting to interpret the last result in terms of a counting statistics. Namely, we might
identify terms $e^{\pm i \chi(t)}$ and $e^{\pm i\chi(t)/2\pm i\chi
  (t')/2}$ with a quasi-charge transfer of $\pm e$ at $t$ and $(\pm
e/2,\pm e/2)$ at $(t,t')$, respectively. The fact, that in this
expression half-integer charges appear has probably a similar origin
as the half-integer charge, which appears in resonant tunneling.
\cite{dejong:96,belzig:03,bagrets:03,pistolesi:04}  Hence, interpreted
as a quasi-Poissonian distribution we may identify "probabilities"
according to
\begin{eqnarray}
  \label{eq:quasipoisson}
  S & \sim & \sum_{\sigma=\pm}p_\sigma(t) e^{i\sigma\chi(t)}+\nonumber\\
  & & \sum_{\sigma,\sigma^\prime}p_{\sigma\sigma^\prime}(t,t^\prime)
  e^{i\sigma\chi(t)/2+i\sigma^\prime\chi(t^\prime)/2 }.
\end{eqnarray}
By comparing Eq.~(\ref{eq:quasipoisson}) to Eq.~(\ref{eq:quasi1}) we can
read off the transfer "probability" per unit time. These take the form
\begin{eqnarray}
p_\sigma(t) & = & \sigma\sum_n T_neV(t)dt/2\pi\hbar,\\
p_{\sigma\sigma^\prime}(t,t') & = & \sigma\sigma^\prime
\sum_n T_ndtdt'\frac{w(t-t')}{2\pi}\cos\varphi(t,t').\nonumber
\end{eqnarray}
Unfortunately, these rates can be negative, so they cannot be interpreted
as a probability. Only a combination of $p_\sigma$ and $p_{\sigma\sigma'}$ would make sense,
which actually happens in usual full counting statistics -- valid at long times.\cite{fcs}
Hence, the generating function (\ref{gent}) itself does not correspond to a measurable
probability. Only after convolution with the Gaussian detection noise we get a
real probability.

\section{Quantum tape}

In section III we have introduced a measurement protocol by means of
Kraus operators.  Here we go further and want to find a quantum
detector that is coupled to a source and registers the time
dependence of the source.  Our aim is make a \emph{quantum tape} that
translates time dependence into spatial dependence. The tape interacts
with the system at a fixed point in space.  Then the tape moves far
from the point of interaction with the source and afterwards the
projection (reading) is applied.

\begin{figure}
\includegraphics[width=0.4\columnwidth]{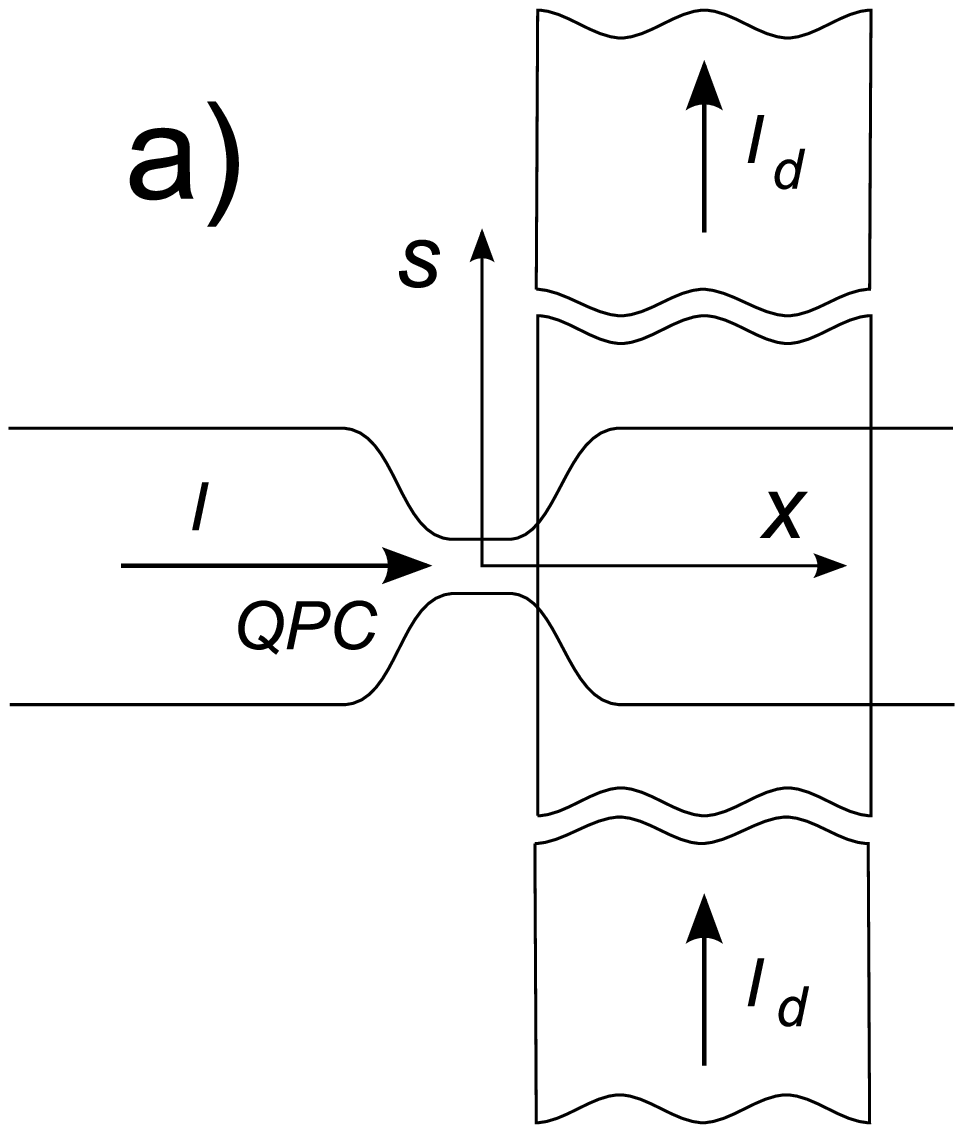}
\includegraphics[width=0.55\columnwidth]{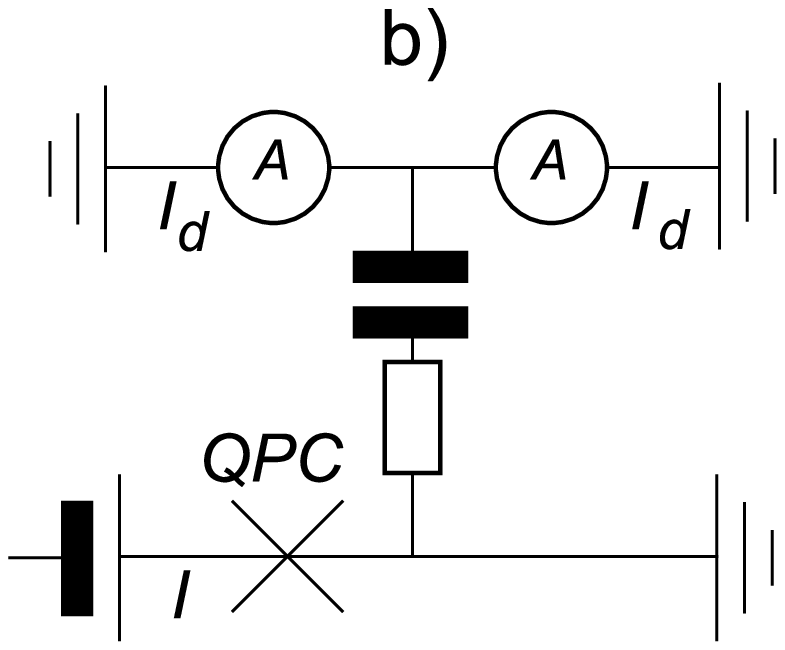}
\caption{The system (quantum point contact) coupled to a one-dimensional wire (a)
and possible equivalent electric circuit (b). Ammeters, switched on at instant time, 
measure spatial profile of the current.}\label{dete}
\end{figure}

Our quantum tape will be a linear quantum wire or equivalently a
massless Josephson transmission line.  In this description we have the
joint Hamiltonian of the detector and the system
\begin{equation}
\hat{H}=\hat{H}_{d}+\hat{H}_{I}+\hat{H}_0\,.\\
\end{equation}
Here
\begin{eqnarray}
&&\hat{H}_{d}=\frac{\pi\hbar}{2e^2}\int ds(\hat{Q}^2_d(s)+\hat{I}^2_d(s)),\nonumber\\
  &&\hat{H}_0=\sum_{\bar n}\int dx\left\{V(t)\theta(-x)\hat{Q}_{\bar n}(x)\right.\label{hamdet}\\
  &&+ i\hbar v_n[\hat{\psi}^\dag_{L\bar
    n}(x) \partial_x\hat{\psi}_{L\bar n}(x)- 
  \hat{\psi}^\dag_{R\bar n}(x)\partial_x\hat{\psi}_{R\bar n}(x)],\nonumber\\
  &&\left.+q_n\delta(x)[\hat{\psi}^\dag_{L\bar n}(x)\hat{\psi}_{R\bar n}(-x)+
    \hat{\psi}^\dag_{R\bar n}(x)\hat{\psi}_{L\bar n}(-x)]\nonumber
  \right\},\\
&&\hat{H}_{I}=-\frac{2\pi\hbar}{e^2}\int ds \lambda(s)\hat{Q}_d(s)\hat{Q}\nonumber
\end{eqnarray}
with $\lambda(s)=\lambda(-s)$, $\hat{Q}=\sum_{\bar n}\int dx f_n(x/v_n)\sum_{\bar n}\hat{Q}_{\bar n}(x)$,
\begin{equation}
\hat{Q}_{\bar n}=\sum_{A=R,L}e\hat{\psi}^\dag_{A\bar n}(x)\hat{\psi}_{A\bar n}(x).
\end{equation}
The bosonic operators satisfy (see also the fermionic representation in appendix F)
\begin{eqnarray}
&&[\hat{I}_d(s),\hat{I}_d(s')]=[\hat{Q}_d(s),\hat{Q}_d(s')]=0\nonumber\\
&&[\hat{I}_d(s),\hat{Q}_d(s')]=-ie^2\partial_s(s-s')/\pi\;.
\end{eqnarray}
We used here time units for the Hilbert space of the detector (Fermi velocity $v_d=1$).
The functions $\lambda$ and $f_n$ define the coupling between the
detector and the system. Their
arguments are in time units and should be nonzero only
on one side of the junction.  In principle it should be the total
charge, i.e.~$f_n(s)=\theta(s)$, but we have to allow finite range of
the coupling. The setup is depicted in Fig.~\ref{dete}.

We also define Heisenberg operators $\hat{A}(t)=\hat{U}^\dag(t)\hat{A}\hat{U}(t)$
with $\hat{U}(t)=\mathcal Te^{\int_0^t\hat{H}(t')dt'/i\hbar}$.
The initial density matrix is $\hat{\rho}=\hat{\rho}_d\hat{\rho}_s$
with $\hat{\rho}_a\propto e^{-\hat{H}_a/k_BT_a}$ for $a=d,0$.
The current operator is defined $\hat{I}=\sum_{\bar n}e\hat{\psi}^\dag_{L\bar n}
\hat{\psi}_{L\bar n}-L\leftrightarrow R$.
We define the auxiliary current operator
\begin{equation}
\hat{\tilde I}(s)=(\hat{I}_d(s)-\hat{I}_d(-s))/2.
\end{equation}
Now, performing a spatially resolved measurement of
$\hat{\tilde{I}}(s,t)$ at $s\gg 0$ (beyond the support of
$\lambda(s)$) we get the probability density functional
\begin{equation}
\rho[\tilde{I}]=\mathrm{Tr}\hat{\rho}\delta[\hat{\tilde I}-\tilde{I}].
\end{equation}
The generating functional $\mathcal S[\chi]=\ln\langle e^{\int
  i\chi(s)\tilde{I}(s-t)ds/e}\rangle$ in the limit of weak coupling
(neglecting terms $\mathcal O[\chi\lambda^2]$) will be given by (see
appendix F)
\begin{equation}
\mathcal S[\chi]
=\tilde{S}[-\dot{\chi}\ast\lambda,i\chi\ast\lambda\ast w_d/2]-\int d\omega|\chi(\omega)|^2w_d(\omega)/16\pi^2\,.\label{weak}
\end{equation}
The convolution in the arguments is defined as $(a\ast\lambda)(\omega)=a(\omega)\lambda(\omega)$,
$\dot{\chi}(\omega)=-i\omega\chi(\omega)$.
and 
\begin{eqnarray}
&&\tilde{S}[\chi,\phi]=\ln\mathrm{Tr}\:\hat{\rho}_0\:\tilde{\mathcal T}\:e^{\int
  idt(\chi(t)+2\phi(t))\hat{Q}_0(t)/2e}\:\times\nonumber\\
&&\mathcal T\:e^{\int idt(\chi(t)-2\phi(t))\hat{Q}_0(t)/2e}.\label{genf02}
\end{eqnarray}
Here we used the interaction picture, in which
$\hat{A}_0(t)=\hat{U}^\dag_0(t)\hat{A}\hat{U}_0(t)$  with
$\hat{U}_0(t)=\mathcal Te^{\int_0^t\hat{H}_0(t')dt'/i\hbar}$.  We get
the expected detection picture -- the measured signal $\tilde
I(\omega)$ is proportional to $\lambda(\omega)I_0(\omega)$ with $\hat{I}_0(t)=d\hat{Q}(t)/dt=
i[\hat{H}_0(t),\hat{Q}_0(t)]/\hbar$. The contribution from $I_0$ is much weaker ($\lambda\ll 1$) than
the internal Gaussian noise of the detector -- the last term in (\ref{weak}).
To evaluate (\ref{genf02}), we can apply the FCS formalism developed
in Sec. III and App. A for $\hat{H}_I=\int dx\sum_{\bar n}\hat{Q}_{\bar
  n}(x)V_{s\bar n}(x,t)$ with the bias $V_{s\bar
  n}(x,t)+V(t)\theta(-x)$,
\begin{equation}
V_{s\bar n}(x,t)=\hbar\phi(t)f_n(x/v_n)/e.
\end{equation} 

The averages measured at the detector are a combination of its own
noise, the noise of the system and the response of the system due to
the detector.  The average
\begin{equation}
\langle \tilde{I}(\omega)\rangle=\sum_n \tilde{g}_n(\omega)T_n e^2V(\omega)/\pi\hbar
\label{current}
\end{equation}
with $\tilde g_n=g_n(1+\mathcal O[g^2])$, $g_n(\omega)=-i\omega\lambda(\omega) f_n(\omega)$.
The observed fluctuation spectral density of $\tilde I$ is
$C_2(\omega,\omega')=\langle\delta \tilde I(\omega)\delta \tilde
I(\omega')\rangle = 2\pi e^2\delta(\omega+\omega')P(\omega) +\sum_n
e^2\tilde{g}_n(\omega)\tilde{g}_n(\omega')P_n(\omega,\omega')$. It contains a
voltage independent contribution
\begin{eqnarray}
&&P(\omega)=w_d(\omega)/4\pi-\sum_nR_nT_n w(\omega)
|g_n(\omega)|^2\label{off}/\pi\label{noo2}\\
&&
+\sum_n[w(\omega)-w_d(\omega)]\left[|g_n(\omega)|^2-R_n\mathrm{Re}\,g_n^2(\omega)\right]/\pi
\,,\nonumber
\end{eqnarray}
where $w(\omega)=\omega\coth(\hbar\omega/2k_BT_0)$. The second part is
\begin{eqnarray}
&&P_n(t,t')=R_nT_n\frac{w(t-t')}{\pi\Gamma_n(t-t')}
\cos\int_t^{t'} du\frac{eV(u)}{\hbar},
\label{exc1}\\
&&
\ln\Gamma_n(t)=\int d\omega\frac{2\sin^2(\omega t)}
{(2\pi\omega)^2}[w_d(\omega)|g_n(\omega)|^2+\mathcal O[g^4]].\nonumber
\end{eqnarray}
The excess noise -- i.e.~the voltage-dependent part of $C_2$ -- is
contained in (\ref{exc1}). Note that this result is valid for an
arbitrary time-dependent bias voltage.  In the limit of weak coupling,
viz.~$g_n\ll 1$, we can take $\Gamma(t)\simeq 1$ and the excess noise
agrees with the symmetrized quantum noise. \cite{blanter,lesovik}  It
is a tiny contribution on top of a huge background noise $P(\omega)$.
The background noise is also seen in experiments but it cannot be
distinguished from the amplifier's noise.  Measurements of the
voltage dependence reveal only the excess noise. \cite{glatt2, gabel}
In the limit of strong coupling, $g_n\gg 1$, the noise is heavily
affected by the detector's backaction. The excess noise is affected through
$\Gamma(t)>1$ and gets a significant correction, similar to that in Sec. III, which vanishes for high voltage since $\Gamma(t\to 0)=1$.

The situation is different for the third cumulant. In the limit of
weak coupling we keep only terms to the lowest order in $g_n$, which
gives
\begin{eqnarray}
&&C_3(\omega_1,\omega_2,\omega_3)=\langle\delta \tilde I(\omega_1)\delta \tilde I(\omega_2)\delta \tilde I(\omega_3)\rangle=\label{tri}\\
&&
\sum_n e^3\prod_l g_n(\omega_l)\left\{\vphantom{\sum_{\sigma(ijk)}}R_nT_n(R_n-T_n)\frac{eV(\omega_1+\omega_2+\omega_3)}{\pi\hbar}\right.\nonumber\\
&&\left.+\sum_{\sigma(ijk)}(2T_n+q_n(\omega_k)-1)P_n(\omega_i,\omega_j;\omega_k)w(\omega_k)/4\right\}\nonumber
\end{eqnarray}
for the susceptibility $P_n(\alpha,\beta;s)$ defined by
(\ref{susdef}),
$q_n(\omega)=e^{i\vartheta_n(\omega)}[1-w_d(\omega)/w(\omega)]$ and
$\vartheta_n(\omega)=-2\arg g_n(\omega)$. The inner summation runs
over all permutations of the set $123$. Note that $q_n$ leads to
corrections that are indistinguishable from the influence of an
environment described by (\ref{encor}) (up to factors $g_n$). The
detection dependent part $g_n(\omega)(q_n(\omega)-1)/\hbar$ from
(\ref{tri}) can be effectively absorbed into $b_0$ in
(\ref{encor}). Hence, the detector can be interpreted as an example of
some environment.

Another interesting observation is that all cumulants, except the
first term in (\ref{noo2}), vanish at equilibrium between detector and
the system, namely for $eV=0$ and $k_BT_0=k_BT_d$.  On the other hand
it is obvious because no information transfer is possible at
equilibrium as the entropy is already maximized. An efficient detector
\emph{cannot} be in equilibrium with the measured system.

\begin{figure}
  \includegraphics[width=0.6\columnwidth,clip=true,angle=270]{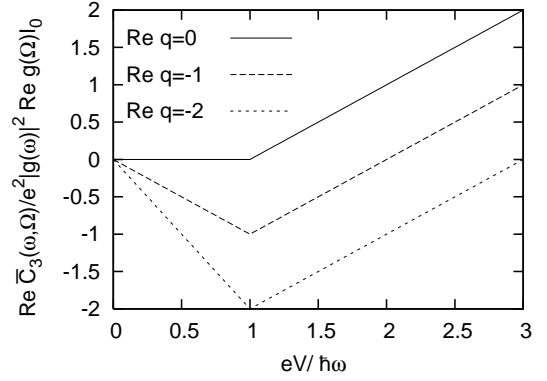}
  \caption{The high-frequency third cumulant of current fluctuations
    in the tunneling limit at zero temperature for different values of $q=q_n$. The average current is
    $I_0=2e^2\sum_n T_nV/h$, we took $|\Omega|\ll(\omega,eV/\hbar)$
    and assumed a mode-independent coupling $g=g_n$. Note that $\mathrm{Re}\,q\leq 0$ at $\vartheta=0$ and $q=0$ at $k_BT_d=0$.}\label{s33}
\end{figure}

Let us finally consider two limiting cases. Analogously to (\ref{sb3}) and (\ref{qzero}), for $k_BT_0=0$ and 
$\Omega\ll(\omega,|eV|/\hbar)$ we get from (\ref{tri})
\begin{eqnarray}
&&\bar{C}_3(\Omega,\omega)
=\sum_n g_n(\Omega)|g_n(\omega)|^22R_nT_n\frac{e^4V}
{h}\times\label{three2}\\
&&
\left[1-2T_n+\left(2T_n-1+\mathrm{Re}\,q_n(\omega)\right)\min\left(1,\frac{\hbar|\omega|}{|eV|}\right)
\right].
\nonumber
\end{eqnarray}
This result is plotted for different $q_n$ in Fig.~\ref{s33} in the
tunneling limit, $T_n\ll 1$. We see, that  $q_n$ induces a step in
$\bar C_3$ for $eV=\hbar\omega$.
For arbitrary $k_BT_0$ and $|\omega_k|\ll |eV|/\hbar$, we have
\begin{eqnarray}
&&\bar{C}_3(\omega_1,\omega_2)
=\frac{e^4V}{h}\sum_n 2R_nT_n\prod_k g_n(\omega_k)\times\\
&&
\left[1-2T_n+\left(2T_n-1+\sum_kq_n(\omega_k)/3\right)(1-F(0,0))\right]\nonumber
\end{eqnarray}
with $F(0,0)$ given by (\ref{qzero1}) in agreement with the environmental correction.
\cite{been} We see, that the behavior of the third cumulant is
different from the noise. Even in the weak coupling limit it may get a
significant quantum correction due to the detector. 

\subsection{Many tapes}

We have seen in Eq.(\ref{weak}) that the outcome of the measurement is always correlated
with fluctuations in the tape as the second argument of $\tilde{S}$ is non-zero.
This can be avoided by means of many tapes, namely

\begin{equation}
  \hat{H} = \hat{H}_0+ \frac{\pi\hbar }{2e^2}
  \sum_{c=1}^N\int [\hat{Q}^2_c(s)+\hat{I}^2_c(s) - 4\lambda_c(s)\hat{Q}_c(s) 
  \hat{Q}] ds.\label{ham2t}
\end{equation}
with 
\begin{eqnarray}
&&[\hat{I}_a(s),\hat{Q}_b(s')] =
-ie^2\delta_{ab}\partial_s\delta(s-s')/\pi,\\
&&[\hat{I}_a(s),\hat{I}_b(s')]=[\hat{Q}_a(s),\hat{Q}_b(s')]=0.\nonumber
\end{eqnarray}
We also redefine
\begin{equation}
\hat{\tilde{I}}(s)=\sum_c(\hat{I}_c(s)-\hat{I}_c(-s)).
\end{equation}
The resulting generating function (\ref{sss}) has the new form
in the limit of weak coupling,
\begin{eqnarray}
&&\mathcal S[\chi]
=\tilde{S}\left[-\sum_c\dot{\chi}\ast\lambda_c,\sum_ci\chi\ast\lambda_c\ast w_c/2\right]
\nonumber\\
&&-\int d\omega|\chi(\omega)/4\pi|^2\sum_cw_c(\omega).\label{weak2}
\end{eqnarray}
The results for the cumulants differ from the single tape case as
follows.  In the definition of $g_n$ below (\ref{current}) we have to
replace $\lambda\to\sum_c\lambda_c$. In (\ref{off}) we change the
first term $w_d\to\sum_c w_c$. We also change
$w_d\to\sum_c\lambda_cw_c/\sum_c\lambda_c$ in the second line of
(\ref{off}) and the definition of $q_n$ below (\ref{tri}).  This
change makes in principle arbitrarily large $q_n$ (positive or negative)
possible even at $\vartheta_n=0$.  On the other hand a simple model of
environment\cite{been} can give effectively $q_n\leq 1$ when comparing
(\ref{tri}) and (\ref{encor}), while the case $q_n>1$ has been
observed in the recent experiment.\cite{regab} The latter suggests that the
actual detection scheme or model of environment is more complicated.

Let us consider a very special case of tapes at high temperatures
$1/\lambda_c\tau\gg k_BT_{c}/\hbar\gg 1/\tau$, where $\tau$ is a
certain time resolution so that the measurable range of frequencies is
$|\tau\omega|\ll 1$.  If $ \sum_c\lambda_ck_BT_c=0$ (but
$\sum_c\lambda_c\neq 0$) then we can write
\begin{equation}
\mathcal S[\chi]
=\tilde{S}[-\sum_c\dot{\chi}\ast\lambda_c,0]
-\int d\omega|\chi(\omega)|^2\sum_cw_c(\omega)/16\pi^2,\label{weak3}
\end{equation}
so that the system-detector correlation cancels ($0$ in the second
argument of $\tilde{S}$). The detector model becomes then
\emph{classical} while still gaining \emph{quantum} information. Such
a situation is close to the POVM model from Sec. III in the limit of weak coupling because
the detector's noise (last term in (\ref{weak3})) becomes white and decouples from the system's signal.

\section{Conclusions}

We have presented detection models of time-resolved quantum detection
of current in mesoscopic junction.  One can make the measurement by
means of either a partial projection (Kraus operator -- POVM) or a
full projection on a weakly coupled transmission line -- quantum
tape. It is difficult to separate the backaction of the detector from
the signal of the system. Nevertheless, the generality of the
presented method makes it applicable to a wide range of types of
measurement.

Applied to a quantum point contact, both models give the expected
expressions for current and noise if the background noise is
subtracted. The voltage-dependent part of the noise is independent of
the detector in the weak coupling limit. On the other hand, the third
cumulant may contain significant voltage-dependent corrections even in
the weak coupling limit. The correction is indistinguishable from the
effect of the environment.  Hence, the high frequency measurement in
mesoscopic junctions always contains a detection-dependent
contribution.  Experimentally, an independent determination of the
coupling parameters would help to distinguish between the effects of
the detection process and the environmental backaction.

We have also derived an analytical tool to calculate
frequency-dependent cumulants for a mesoscopic junction with
energy-independent transmission. An important result is to identify
the noise susceptibility in the expression for the third cumulant at
high frequency.  The method works well both in time and frequency
domain and can be extended to other complicated correlations.

\section*{Acknowledgments} 
We are grateful to J. Gabelli and B. Reulet for discussions.
We acknowledge  financial support from the German Research Foundation
(DFG) through SFB 767 and SP 1285
\textit{Semiconductor Spintronics}.

\section*{Appendix A}
\renewcommand{\theequation}{A\arabic{equation}}
\setcounter{equation}{0}

In our model of the quantum point contact, we make the approximation
of a linear spectrum and a short range scattering potential. Both
assumptions lead to singular behaviour and we have to regularize some
terms in the Hamiltonian and describe the details here.

The short range potential scattering is introduced in Eq.~
(\ref{ham2}) and we would like to relate this term to the scattering matrix.
One of the possible regularizations of (\ref{ham2}) is
\begin{equation}
\sum_{\bar n}\int dx z'_n(x)v_n\left[\hat{\psi}^\dag_{L\bar
    n}(x)\hat{\psi}_{R\bar n}(x)+\hat{\psi}^\dag_{R\bar
    n}(x)\hat{\psi}_{L\bar n}(x)\right]
\end{equation}
with the function $z_n(x)\to q_n\theta(x)$. For $|k_nx|\ll 1$, the wave eigenfunctions satisfy
\ba
&&i\hbar\partial_x\psi_{L\bar n}(x)+z'_n(x)\psi_{R\bar n}(x)=0,\nonumber\\
&&-i\hbar\partial_x\psi_{R\bar n}(x)+z'_n(x)\psi_{L\bar n}(x)=0.
\ea
We can change variables $x\to z(x)$,
\ba
&&i\hbar\partial_{z_n}\psi_{L\bar n}+\psi_{R\bar n}=0,\nonumber\\
&&-i\hbar\partial_{z_n}\psi_{R\bar n}+\psi_{L\bar n}=0,
\ea
which gives the solution 
\ba
&&
\psi_{L\bar n}=c_+e^{z_n/\hbar}+c_-e^{-z_n/\hbar},\nonumber\\
&&
\psi_{R\bar n}=-ic_+e^{z_n/\hbar}+ic_-e^{-z_n/\hbar}.
\ea
By taking $\psi_{L\bar n}(z_n=q_n)=0$,$\psi_{R\bar n}(z_n=0)=1$, we get
\ba
&&\tilde{t}_n=\psi_{R\bar n}(z_n=q_n)=\frac{1}{\cosh(q_n/\hbar)},\nonumber\\
&&\tilde{r}_n=\psi_{L\bar n}(z_n=0)=-i\tanh(q_n/\hbar).
\ea
The phase factor in $\tilde{t}_n$
is completely irrelevant in our approximation.
However, it can be added by a constant, symmetric bias potential. 

For the future convenience we shall use a little different regularization
of the second term in (\ref{ham2}), namely
\begin{equation}
\sum_{\bar n}\int dx z'_n(x/v_n)\left[\hat{\psi}^\dag_{L\bar
    n}(x)\hat{\psi}_{R\bar n}(-x)+\hat{\psi}^\dag_{R\bar
    n}(-x)\hat{\psi}_{L\bar n}(x)\right].
\end{equation}
It gives $\tilde{t}_n=\cos(q_n/\hbar)$ and $\tilde{r}_n=i\sin(q_n/\hbar)$.
The regularized Hamiltonian (\ref{hscat}) takes the form
\ba
&&\hat{H}=\hat{H}_0+\sum_{\bar n}\int ds\;2z'_n(s)\hat{I}_{3\bar n}(s)/e\nonumber\\
&&+\sum_{\bar n,\pm}\int ds\;V(\pm sv_n,t)[\hat{I}_{0\bar n}(s)\pm\hat{I}_{1\bar n}(s)]
\ea
To derive (\ref{comm}) and (\ref{comh}), we recall the standard
bosonization scheme for a 1D system of noninteracting
fermions.\cite{lut} Let us define
\begin{equation}
\hat{\psi}_{L\bar n}(k)=\int\frac{dx}{\sqrt{2\pi}}e^{ikx}\hat{\psi}_{L\bar n}(x).
\end{equation}
We assume that the excitations of the Fermi sea are finite in the sense that 
levels deep below and high above the Fermi level are always occupied and empty, respectively.
This leads to
\begin{equation}
\hat{\psi}_{L\bar n}(k\to -\infty),\hat{\psi}^\dag_{L\bar n}(k\to \infty) \to 0,
\label{asym}
\end{equation}
when acting on the density matrix standing to the \emph{right} of these operators.
Next we construct the operators
\ba
&&\hat{A}(x)=\hat{\psi}^\dag_{L\bar n}(x) \hat{\psi}_{L\bar n}(x)=
\int\frac{dk}{\sqrt{2\pi}}e^{-ikx}\hat{A}(k),\nonumber\\
&&\hat{A}(k)=
\int \frac{dq}{\sqrt{2\pi}}\hat{\psi}^\dag_{L\bar n}(q)\hat{\psi}_{L\bar n}(q+k).
\ea
The last operator needs to be regularized as 
\begin{equation}
\hat{A}(k)=
\int \frac{dq}{\sqrt{2\pi}}g(q)\hat{\psi}^\dag_{L\bar n}(q)\hat{\psi}_{L\bar n}(q+k),
\end{equation}
where $g(q)=1$ for $|q|\ll\Lambda$ and $g(q)\to 0$ for $q\gg \Lambda$, with
$\Lambda$ denoting some cutoff larger than all relevant wave scales.
Then
\begin{equation}
\{\hat{\psi}^\dag_{L\bar n}(q),\hat{\psi}_{L\bar n}(k)\}=\delta(q-k).
\end{equation}
Hence,
\ba
&&[\hat{A}(k),\hat{A}(k')]=\int\frac{dq}{2\pi}\hat{\psi}^\dag_{L\bar n}(q+k+k')
\hat{\psi}_{L\bar n}(q)\times\nonumber\\
&&g(q)(g(q+k')-g(q+k)).
\ea
For $|q|,|k|,|k'|\ll\Lambda$ the above expression gives zero because
$g(q+k')=g(q+k)$. Also for $|q|\gg\Lambda$, it zero due to $q(q)=0$.
For $q\sim -\Lambda$ we get again zero due to (\ref{asym}).
The only nonzero terms come from $q\sim \Lambda$. We can use here also (\ref{asym})
but only after changing the order of the operators $\hat{\psi}$
and $\hat{\psi}^\dag$. The net contribution is the anticommutator for $q>0$.
Hence,
\ba
&&[\hat{A}(k),\hat{A}(k')]=\\
&&
\int_{0}^{\infty}\frac{dq}{2\pi}\delta(k+k')g(q)(g(q+k')-g(q+k))=\nonumber\\
&&=\delta(k+k')\int_{0}^{\infty}\frac{dq}{2\pi}g(q)(g(q-k)-g(q+k))=\nonumber\\
&&\delta(k+k')\int_0^k\frac{dq}{2\pi}
g(q)g(q-k)=k\delta(k+k')/2\pi.\nonumber
\ea
In the last line we used the fact that $g(q)=1$ for $q,k\ll\Lambda$.
By taking the Fourier transform we get
\begin{equation}
[\hat{A}(x),\hat{A}(x')]=i\partial_x\delta(x-x')/2\pi.\label{boss}
\end{equation}

\section*{Appendix B}
\renewcommand{\theequation}{B\arabic{equation}}
\setcounter{equation}{0}

The current operators defined in (\ref{ijn}) can be found exactly by a
solution of their respective equations of motion.  We have the
following Heisenberg equations (\ref{heis}) for (\ref{ijn}), 
\ba
&&D\hat{I}_{0\bar n}(s,t)=e^2\partial_sV_+(sv_n,t)/4\pi,\label{hei}\\
&&D\hat{I}_{1\bar n}(s,t)=-2z'_n(s)\hat{I}_{2\bar n}(s,t)+
e^2\partial_s V_-(sv_n,t)/4\pi,\nonumber\\
&&D\hat{I}_{2\bar n}(s,t)=2z'_n(s)\hat{I}_{1\bar
  n}(s,t)-eV_-(sv_n,t)\hat{I}_{3\bar n}
(s,t),\nonumber\\
&&D\hat{I}_{3\bar n}(s,t)=eV_-(sv_n,t)\hat{I}_{2\bar n}
(s,t),\nonumber 
\ea 
where $D=\hbar(\partial_t-\partial_s)$ and
$V_\pm(x,t)=V(x,t)\pm V(-x,t)$. For $s\to+\infty$ we have asymptotic
equilibrium operators
\begin{equation}
\hat{I}_{j\bar n}(s,t)\to\hat{I}_{j\bar n}(t_s)\label{eqq}
\end{equation}
for $j=0,1,2,3$ and $t_s=t+s$.
The operator $\hat{I}_0$ decouples from the other currents and has the general solution
\begin{equation}
\hat{I}_{0\bar n}(s,t)=\hat{I}_{0\bar n}(t_s )+\frac{e^2}{4\pi\hbar}\partial_s\int_s^\infty
ds'V_+(s'v_n,t+s-s').
\end{equation}
For $s>0$ we have
\ba
&&\hat{I}_{1\bar n}(s,t)=\hat{I}_{1\bar n}(t_s)+\frac{e}{4\pi}\partial_s\Phi_n(s,t),\nonumber\\
&&\hat{I}_{2\bar n}(s,t)=\hat{I}_{2\bar n}(t_s)\cos\Phi_n(s,t)
-\hat{I}_{3\bar n}(t_s)\sin\Phi_n(s,t),\nonumber\\
&&\hat{I}_{3\bar n}(s,t)=\hat{I}_{3\bar n}(t_s)\cos\Phi_n(s,t)
+\hat{I}_{2\bar n}(t_s)\sin\Phi_n(s,t),\nonumber\\
&&
\Phi_n(s,t)=\frac{e}{\hbar}\int_s^\infty
ds'V_-(s'v_n,t+s-s')
\ea
and for $s<0$
\ba
&&\hat{I}_{1\bar n}(s,t)=\hat{I}^0_{1\bar n}(t_s)+\frac{e}{4\pi}\partial_s\Phi_{+n}(s,t),\nonumber\\
&&\hat{I}_{2\bar n}(s,t)=\hat{I}^0_{2\bar n}(t_s)\cos\Phi_{+n}(s,t)
-\hat{I}^0_{3\bar n}(t_s)\sin\Phi_{+n}(s,t),\nonumber\\
&&\hat{I}_{3\bar n}(s,t)=\hat{I}^0_{3\bar n}(t_s)\cos\Phi_{+n}(s,t)
+\hat{I}^0_{2\bar n}(t_s)\sin\Phi_{+n}(s,t),\nonumber\\
&&
\Phi_{+n}(s,t)=\Phi_n(s,t)-\Phi_n(t_s),\:\Phi_n(t)=\Phi_n(0,t),
\ea
with the boundary conditions
\ba
&&\hat{I}^0_{1\bar n}(s)=(T_n-R_n)\hat{I}_{1\bar n}(0_+,s)
-2\sqrt{R_nT_n}\hat{I}_{2\bar n}(0_+,s),\nonumber\\
&&\hat{I}^0_{2\bar n}(s)=(T_n-R_n)\hat{I}_{2\bar n}(0_+,s)
+2\sqrt{R_nT_n}\hat{I}_{1\bar n}(0_+,s),\nonumber\\
&&\hat{I}^0_{3\bar n}(s)=\hat{I}_3(0_+,s).\nonumber
\ea
In particular for $s<0$
\ba
&&\hat{I}_{1\bar n}(s,t)=(T_n-R_n)\hat{I}_{1\bar n}(t_s)\\
&&+\frac{e}{4\pi}\partial_s(\Phi_n(s,t)-2R_n\Phi_n(t_s))
-2\sqrt{R_nT_n}\hat{J}^{\Phi}_{\bar n}(t_s)\nonumber
\ea
with
\begin{equation}
\hat{J}^{\Phi}_{\bar n}(s)=\hat{I}_{2\bar n}(s)\cos\Phi_n(s)-
\hat{I}_{3\bar n}(s)\sin\Phi_n(s).\label{defj2}
\end{equation}
The great advantage of the above equations is that the Heisenberg current operators
can be represented by combinations of equilibrium operators. The same
applies to averages and facilitates the calculation of higher cumulants.

\section*{Appendix C}
\renewcommand{\theequation}{C\arabic{equation}}
\setcounter{equation}{0}

To evaluate the noise correction due to the detetctor in the POVM
model we employ the the formalism developed previously. We have to
perform a number of path integrals to find the final analytical
expressions for the noise. We start by splitting Eq.~(\ref{irr}) into
two parts 
\ba
&&\hat{I}_R(t)=\sum_{\bar n}\left[\hat{I}^{\mathrm{in}}_{R\bar n}(t)+\hat{I}^{\mathrm{out}}_{R\bar n}(t)\right],\\
&&\hat{I}^{\mathrm{out}}_{R\bar n}(t)=\int dx\; g(x+x_R)\left[\hat{I}_{0\bar n}(x_n,t)-\hat{I}_{1\bar n}(x_n,t)\right],\nonumber\\
&&\hat{I}^{\mathrm{in}}_{R\bar n}(t)=\int dx\;
g(x-x_R)\left[-\hat{I}_{0\bar n}(x_n,t)-\hat{I}_{1\bar
    n}(x_n,t)\right],\nonumber 
\ea 
with $x_n=x/v_n$. To simplify
(\ref{genf1}), we will use the commutators
\ba
&&[\hat{I}^{\mathrm{in}}_{R\bar n}(t),\hat{I}^{\mathrm{out}}_{R\bar
  n}(t')]=0\mbox{ for }
t>t'\\
&&[\hat{I}^{\mathrm{in}}_{R\bar n}(t),\hat{I}^{\mathrm{in}}_{R\bar
  n}(t')]= [\hat{I}^{\mathrm{out}}_{R\bar
  n}(t),\hat{I}^{\mathrm{out}}_{R\bar n}(t')]=
\frac{ie^2}{2\pi}h'_n(t-t').\nonumber \ea 
with $h_n$ defined by
(\ref{tndef}). Using the Baker-Hausdorff formula
$e^{\hat{A}}e^{\hat{B}}=e^{\hat{A}+\hat{B}}e^{[\hat{A},\hat{B}]/2}$ we
find 
\ba
 &&\mathcal Te^{\int \frac{idt}{2e}\xi(t)\hat{I}_R(t)}=
\prod_{\bar n}\exp\int\frac{idt}{2e}\xi(t)\hat{I}^{\mathrm{out}}_{R\bar n}(t)\;\times\\
&&\exp\int\frac{idt}{2e}\xi(t)\hat{I}^{\mathrm{in}}_{R\bar
  n}(t)\exp\int\frac{-idtdt'}{8\pi}\xi(t)\xi(t')
\theta_n(t-t')\,.\nonumber 
\ea 
Using similar transformations we get
\begin{equation}
e^{S[\chi,\phi]}=\mathrm{Tr}\hat{\rho}\prod_{\bar n}\hat{U}^{\mathrm{in}}_{\bar n}
\hat{U}^{\mathrm{out}}_{\bar n}\hat{U}^{\mathrm{in}}_{\bar n}e^{i\vartheta_{n}},
\label{gene}
\end{equation}
where
\ba
&&\hat{U}^{\mathrm{in}}_{\bar n}=\exp\int\frac{idt}{2e}\chi(t)\hat{I}^{\mathrm{in}}_{R\bar n}(t),\nonumber\\
&&\hat{U}^{\mathrm{out}}_{\bar n}=\exp\int\frac{idt}{e}\chi(t)\hat{I}^{\phi}_{R\bar n}(t),\nonumber\\
&&\hat{I}^{\phi}_{R\bar n}(t)=\hat{U}_{\phi\bar n}\hat{I}^{\mathrm{out}}_{R\bar n}(t)
\hat{U}^\dag_{\phi\bar n},\nonumber\\
&&\hat{U}_{\phi\bar n}=\exp\int\frac{idt}{e}\phi(t)\hat{I}^{\mathrm{in}}_{R\bar n}(t),\nonumber\\
&&\vartheta_{n}=
\int\frac{dtdt'}{\pi}\chi(t)\phi(t')\theta_n(t-t')
\ea
with $\theta_n$ defined by (\ref{tndef}). Next, to find the noise (\ref{pnoise}), we need the help of the relation
\begin{equation}
\int D\phi \exp\left(-\int dt\left[\frac{2\phi^2(t)}{\tau}+i\phi(t)\xi(t)\right]\right)=
e^{-\int \frac{\tau dt}{8}\xi^2(t)},\label{phiav}
\end{equation}
which gives
\ba
&&P(a,b)=\delta(a-b)/\tau+
\nonumber\\
&&\int dt\sum_{nm}\frac{\tau}{\pi^2}\theta_n(a-t)\theta_m(b-t)
+\int D\phi\; e^{-\int \frac{2dt}{\tau}\phi^2(t)}\nonumber\\
&&\left\{\left[\frac{1}{2}\sum_{\bar n}\mathrm{Tr}\hat{\rho}(\delta\hat{I}^{\mathrm{in}}_{R\bar n}(a)+
\delta\hat{I}^\phi_{R\bar n}(a))(\delta\hat{I}^{\mathrm{in}}_{R\bar n}(b)+
\delta\hat{I}^\phi_{R\bar n}(b))\right.\right.\nonumber\\
&&\left.\left.+\sum_{n\bar m}
\int\frac{2dt}{\pi}\phi(t)\theta_n(a-t)
\mathrm{Tr}\hat{\rho}\hat{I}^\phi_{R\bar m}(b)\right]+a\leftrightarrow b\right\}\nonumber\\
\ea
and
$\delta\hat{A}=\hat{A}-\mathrm{Tr}\hat{\rho}\hat{A}$.
Finally, we get
\ba
&&\delta\hat{I}^{\mathrm{in}}_{R\bar n}(t)=-\hat{I}_{0\bar n}(t_{+n})
-\hat{I}_{1\bar n}(t_{+n}),\nonumber\\
&&\delta\hat{I}^{\phi}_{R\bar n}(t)=-\frac{eR_n}{2\pi}\phi'(t-2t_n)+\nonumber\\
&&\hat{I}_{0\bar n}(t_{-n})
-(T_n-R_n)\hat{I}_{1\bar n}(t_{-n})+2\sqrt{R_nT_n}\hat{J}^{\phi}_{\bar n}(t_{-n}),\nonumber\\
&&\phi_n(t)=\Phi(t)+\int \sqrt{2}dt'\phi(t_{-n}+t')h_n(\sqrt{2}t')
\ea
where $\hat{J}$ is defined by (\ref{defj2}) with $\Phi_n$ replaced
by $\phi_n$.

The case of many detectors (Eqs. (\ref{krau2})-(\ref{mlast})) is analogous,
\ba
&&\hat{I}_A(t)=\sum_{\bar n}\left[\hat{I}^{\mathrm{in}}_{A\bar n}(t)+\hat{I}^{\mathrm{out}}_{A\bar n}(t)\right],\\
&&\hat{I}^{\mathrm{out}}_{A\bar n}(t)=\!\!\int dxg(x+|x_A|)\!\left[\epsilon_A\hat{I}_{0\bar n}(x_n,t)-\hat{I}_{1\bar n}(x_n,t)\right],\nonumber\\
&&\hat{I}^{\mathrm{in}}_{A\bar n}(t)=\!\!\int dxg(x-|x_A|)\!\left[-\epsilon_A\hat{I}_{0\bar n}(x_n,t)-\hat{I}_{1\bar n}(x_n,t)\right]\nonumber
\ea
with $\epsilon_A=\mathrm{sgn}\; x_A$. In Eq. (\ref{gene}) we put
\ba
&&\hat{U}^{\mathrm{in}}_{\bar n}=\exp\int\frac{idt}{2e}\sum_A\chi_A(t)\hat{I}^{\mathrm{in}}_{A\bar n}(t),\nonumber\\
&&\hat{U}^{\mathrm{out}}_{\bar n}=\exp\int\frac{idt}{e}\sum_A\chi_A(t)\hat{I}^{\phi}_{A\bar n}(t),\nonumber\\
&&\hat{I}^{\phi}_{A\bar n}(t)=\hat{U}_{\phi\bar n}\hat{I}^{\mathrm{out}}_{A\bar n}(t)
\hat{U}^\dag_{\phi\bar n},\nonumber\\
&&\hat{U}_{\phi\bar n}=\exp\int\frac{idt}{e}\sum_A\phi_A(t)\hat{I}^{\mathrm{in}}_{A\bar n}(t),\nonumber\\
&&\vartheta_{n}=
\int\frac{dtdt'}{\pi}\sum_A\chi_A(t)\phi_A(t')\theta_{An}(t-t').
\ea
We finally obtain
\ba
&&P_{AB}(a,b)=\delta(a-b)\delta_{AB}/\tau_A+
\nonumber\\
&&\int dt\sum_{nm}\frac{\tau_A\delta_{AB}}{\pi^2}\theta_{An}(a-t)\theta_{Am}(b-t)\nonumber\\
&&
+\int D\phi\; e^{-\int \sum_C\frac{2dt}{\tau_C}\phi^2_C(t)}\nonumber\\
&&\left\{\left[\frac{1}{2}\sum_{\bar n}\mathrm{Tr}\hat{\rho}(\delta\hat{I}^{\mathrm{in}}_{A\bar n}(a)+
\delta\hat{I}^\phi_{A\bar n}(a))(\delta\hat{I}^{\mathrm{in}}_{B\bar n}(b)+
\delta\hat{I}^\phi_{B\bar n}(b))\right.\right.\nonumber\\
&&\left.\left.+\sum_{n\bar m}
\int\frac{2dt}{\pi}\phi_A(t)\theta_{An}(a-t)
\mathrm{Tr}\hat{\rho}\hat{I}^\phi_{B\bar m}(b)\right]\right.\nonumber\\
&&
\left.\vphantom{\sum_{\bar n}}+Aa\leftrightarrow Bb\right\}\nonumber\\
\ea
and
\ba
&&\delta\hat{I}^{\mathrm{in}}_{A\bar n}(t)=-\epsilon_A\hat{I}_{0\bar n}(t)
-\hat{I}_{1\bar n}(t),\nonumber\\
&&\delta\hat{I}^{\phi}_{A\bar n}(t)=\frac{e}{2\pi}(T_n\phi'_B(t)-R_n\phi'_A(t))+\nonumber\\
&&\epsilon_A\hat{I}_{0\bar n}(t)
-(T_n-R_n)\hat{I}_{1\bar n}(t)+2\sqrt{R_nT_n}\hat{J}^{\phi}_{1\bar n}(t),\nonumber\\
&&\phi_n(t)=\Phi(t)+\int \sqrt{2}dt'\sum_A\phi_A(t+t')h_{An}(\sqrt{2}t'),\nonumber
\ea
with $B=L,R$ for $A=R,L$, respectively.

\section*{Appendix D}
\renewcommand{\theequation}{D\arabic{equation}}
\setcounter{equation}{0}
The third cumulant (\ref{third2}) has the form
\begin{equation}
\langle\langle I(a)I(b)I(c)\rangle\rangle=\sum_{\bar n}\mathrm{Tr}\hat{\rho}\hat{F}_{\bar n}(a,b,c)\,.
\end{equation}
The operator in the last equation has the form
\ba
&&
\hat{F}_{\bar n}(a,b,c)
=\sum_{\sigma(abc)}\left[\frac{1}{6}\hat{I}^\Phi_{\bar n}(a_{-n})
\hat{I}^\Phi_{\bar n}(b_{-n})\hat{I}^\Phi_{\bar n}(c_{-n})\right.\nonumber\\
&&
-\frac{1}{4}\left(\hat{I}_{1\bar n}(a_{+n})
\hat{I}^\Phi_{\bar n}(b_{-n})\hat{I}^\Phi_{\bar n}(c_{-n})\right.\\
&&\left.\left.
+\hat{I}^\Phi_{\bar n}(a_{-n})
\hat{I}^\Phi_{\bar n}(b_{-n})\hat{I}_{1\bar n}(c_{+n})\right)\right]\,.\nonumber
\ea
Here a summation over all permutations of the set $abc$ is assumed and 
\ba
&&\hat{I}^\Phi_{\bar n}(t)=
(R_n-T_n)\hat{I}_{1\bar n}(t)+2\sqrt{R_nT_n}\hat{J}^{\Phi}_{\bar n}(t).
\ea
In these equations $\hat{J}$ is defined by (\ref{defj}) and $\Phi$ by (\ref{defphi}).
Using the commutation rules (\ref{comm}) and (\ref{ther}) we get \cite{zaikin1,salo}
\begin{eqnarray}
&&\langle\langle I(a)I(b)I(c)\rangle\rangle=
\sum_{n}R_nT_ne^3\left[\sum_{\sigma(abc)}\sin\varphi(a_{-n},b_{-n})\right.
\nonumber\\
&&[(2T_n-1)q(a_{-n},c_{-n},b_{-n})+q(a_{-n},c_{+n},b_{-n})]/2+\nonumber\\
&&\left.(R_n-T_n)\frac{eV(a_{-n})}{\pi\hbar}\delta(a-b)\delta(a-c)\right]
\label{third}.
\end{eqnarray}
The Fourier transform of $q(a,c,b)$ has the form
\begin{equation}
q(\alpha,\gamma,\beta)=i\delta(\alpha+\beta+\gamma)u(\gamma)(w(\alpha)-w(\beta)),
\end{equation}
where $u$ and $w$ are defined by (\ref{defuw}). The function $\varphi$ is defined by Eq. (\ref{defvphi}).
One can also find that\cite{zaikin1}
\begin{eqnarray}
&&q(a,c,b)=w(b-a)\int_b^a ds\; w(s-c)=\nonumber\\
&&\mathrm{Re}\frac{\sinh(B-A)}{
\sinh^2(B-A+i\epsilon)}\times\nonumber\\
&&\mathrm{Re}\frac{(k_BT/\hbar)^3}{2\sinh(C-A-i\epsilon)
\sinh(B-C+i\epsilon)}\nonumber
\end{eqnarray}
for $X=\pi k_BTx/\hbar$, $x=a,b,c$ and $\epsilon\to 0$. Special care must be taken at 
$a=b$, $b=c$, $c=a$. Note, that the function $q(a,c,b)$ is not cyclic because
\begin{equation}
q(b,a,c)-q(a,c,b)=\partial_b[\delta(a-b)\delta(c-b)]/2\pi.
\end{equation}

\section*{Appendix E}
\renewcommand{\theequation}{E\arabic{equation}}
\setcounter{equation}{0}

In the limiting cases ($R_n\ll 1$ or $T_n\ll 1$), it is possible to
find expressions for the generating functional (\ref{san}). Here we
present some details on the derivation.
The following operator expansion formula is useful,
\begin{eqnarray}
&&e^{\hat{A}+\hat{B}}=e^{\hat{A}}+\int_0^1dx\;e^{x\hat{A}}\hat{B}e^{(1-x)\hat{A}}
\label{expd}\\
&&
+\int_0^1dx\int_0^{1-x}dy\;e^{x\hat{A}}\hat{B}e^{(1-x-y)\hat{A}}\hat{B}e^{y\hat{A}}+\dots\;.
\nonumber
\end{eqnarray}
We use it for $\hat{A}=(R_n-T_n)\int dt\;i\chi(t_{+n})\hat{I}_{1\bar n}(t)/e$ and
\begin{equation}
\hat{B}=2\sqrt{R_nT_n}\int dt\;i\chi(t_{+n})\hat{J}^\Phi_{\bar n}(t)/e.
\end{equation}
What we need is the term of the second power in $\hat{B}$ and
an algebraic identity
\begin{eqnarray}
&&\int_0^1 dx\int_0^{1-y} dy\:\xi\eta\exp[i(x-1/2)\xi+i(y-1/2)\eta]=\nonumber\\
&&2\sin\frac{\xi}{2}\sin\frac{\eta}{2}+i\sin\frac{\xi+\eta}{2}-
i\frac{\xi+\eta}{\xi-\eta}\sin\frac{\xi-\eta}{2}.
\end{eqnarray}
We also need several auxiliary operator identities. Let us define
\begin{equation}
\hat{E}_{\bar n}[\xi]=\exp\left(-\int idt\;\xi(t)
\hat{I}_{1\bar n}(t)/e\right)
\end{equation}
and
\begin{equation}
S_0[\xi]=\ln\mathrm{Tr}\hat{\rho}\hat{E}_{\bar n}^2[\xi]
\end{equation}
for an arbitrary function $\xi$. One can show using (\ref{comm}) and (\ref{ther}) that
\begin{equation}
\int dt\;e^{i\omega t}\frac{\delta\mathrm{Tr}\hat{\rho}\hat{E}^2_{\bar n}[\xi]}{\delta\xi(t)}
=iu(\omega)\mathrm{Tr}\hat{\rho}[\hat{E}^2_{\bar n}[\xi],\hat{I}_{1\bar n}(\omega)]
\end{equation}
and
\begin{equation}
[\hat{E}^2_{\bar n}[\xi],\hat{I}_{1\bar n}(\omega)]=\hat{E}^2_{\bar n}[\xi]
\int\frac{i\omega dt}{2\pi}e^{i\omega t}\xi(t).
\end{equation}
So
\begin{equation}
\int dt\;e^{i\omega t}\frac{\delta S_{0}[\xi]}{\delta\xi(t)}=
-\frac{w(\omega)}{2\pi}\int dt\;e^{i\omega t}\xi(t)
\end{equation}
and we finally obtain (\ref{s00}). Another useful property is
\begin{equation}
\hat{I}_{\pm\bar n}(t)\hat{E}_{\bar n}[\xi]=\hat{E}_{\bar n}[\xi]\hat{I}_{\pm\bar n}(t)
e^{\pm i\xi(t)}.
\end{equation}
We define the operator
\begin{equation}
\hat{D}_{\bar n}(a,b)[\xi]=
\hat{E}_{\bar n}[\xi]\hat{I}_{+\bar n}(a)
\hat{I}_{-\bar n}(b)\hat{E}_{\bar n}[\xi]
\end{equation}
with $\hat{I}_{\pm\bar n}(t)=\hat{I}_{2\bar n}(t)
\pm i\hat{I}_{3\bar n}(t)$.
We can show that
\begin{equation}
\int \!\!dt\;e^{i\omega t}\frac{\delta
\mathrm{Tr}\hat{\rho}\hat{D}_{\bar n}(a,b)[\xi]}{\delta\xi(t)}
=iu(\omega)\mathrm{Tr}\hat{\rho}[
\hat{D}_{\bar n}(a,b)[\xi],\hat{I}_{1\bar n}(\omega)]
\end{equation}
and
\begin{eqnarray}
&&[
\hat{D}_{\bar n}(a,b)[\xi],\hat{I}_{1\bar n}(\omega)]=\hat{D}_{\bar n}(a,b)[\xi]\nonumber\\
&&\left(e^{i\omega b}-e^{i\omega a}+
\int\frac{i\omega dt}{2\pi}e^{i\omega t}\xi(t)\right).
\end{eqnarray}
This gives
\begin{equation}
\ln\frac{\mathrm{Tr}\hat{\rho}\hat{D}_{\bar n}(a,b)[\xi]}
{\mathrm{Tr}\hat{\rho}\hat{D}_{\bar n}(a,b)[0]}
=S_{0}[\xi]+D(a,b)[\xi],
\end{equation}
where $D$ was defined in Eq.~(\ref{ddd}). Finally, we get
\begin{equation}
S[\chi,0]=\sum_n 2(S_0[\chi_{0n}]+S_0[\chi_{1n}])+2\sum_n S_n[\chi],
\end{equation}
where $2\chi_{0n}(t)=\chi(t_{-n})-\chi(t_{+n})$
and $2\chi_{1n}(t)=\chi(t_{-n})+(T_n-R_n)\chi(t_{+n})$.
The lowest order correction is
\begin{eqnarray}
&&S_n[\chi]=\int dt\left\{
R_nT_n\frac{\delta S_0[\chi_{1n}]}{\delta\chi_{1n}(t)}\times\right.\nonumber\\
&&
\left[\frac{\chi(t_{+n})}{T_n-R_n}-\frac{\sin((T_n-R_n)\chi(t_{+n}))}{(T_n-R_n)^2}\right]+\frac{ieV(t)}{2\pi\hbar}\times\nonumber\\
&&\left.
\left[
\frac{T^2_n}{T_n-R_n}\chi(t_{+n})
-R_nT_n\frac{\sin((T_n-R_n)\chi(t_{+n}))}{(T_n-R_n)^2}
\right]
\right\}\nonumber\\
&&-\frac{R_nT_n}{(T_n-R_n)^2}\int dtdt'
\frac{w(t-t')}{\pi}e^{i\varphi(t',t)+D(t',t)[\chi_{1n}]}\times\nonumber\\
&&
\sin\frac{(T_n-R_n)\chi(t_{+n})}{2}
\sin\frac{(T_n-R_n)\chi(t'_{+n})}{2}.
\end{eqnarray}
We stress that the above formula are exact up to first order in
$T_n\ll 1$
or $R_n\ll 1$. After taking the respective limit higher order
contributions should be disregarded as we would need also
$\hat{B}^4$-terms. 

Note also that $\hat{I}_{o\bar n}=(R_n-T_n)\hat{I}_{1\bar{n}}+2\sqrt{R_nT_n}
\hat{I}_{2\bar n}$ satisfies the same commutation rules with itself
and the Hamiltonian 
as $\hat{I}_{1\bar n}$. This leads to the identity
\begin{eqnarray}
&&0=\int dtdt'\; w(t-t')\times\\
&&\left\{
\chi(t')\sin\chi(t)
-4\sin\frac{\chi(t)}{2}\sin
\frac{\chi(t')}{2}e^{D(t,t')[\chi]}\right\}\nonumber
\end{eqnarray}
 for a sufficiently regular function $\chi$.

\section*{Appendix F}
\renewcommand{\theequation}{F\arabic{equation}}
\setcounter{equation}{0}

In the bosonized version (appendix A), we can write in (\ref{hamdet})
\begin{eqnarray}
&&\hat{H}_{d}=
\int dsi\hbar \left[\hat{\psi}^\dag_{l}(s)\partial_s\hat{\psi}_{l}(s)-\hat{\psi}^\dag_{r}(s)\partial_s
\hat{\psi}_{r}(s)\right],\nonumber\\
&&\hat{I}_d(s) =e
(\hat{\psi}^\dag_{r}(s) \hat{\psi}_{r}(s)
-\hat{\psi}^\dag_{l}(s) \hat{\psi}_{l}(s)),\\
&&\hat{Q}_d=e(\hat{\psi}^\dag_r(s)\hat{\psi}_r(s)+\hat{\psi}^\dag_{l}(s) \hat{\psi}_{l}(s))
\nonumber
\end{eqnarray}
with
$\{\hat{\psi}_{a}(s),\hat{\psi}^\dag_{b}(s')\}=\delta_{ab}\delta(s-s')$
for $a,b=l,r$. To derive (\ref{weak}), we first introduce the auxiliary decomposition
\begin{equation}
\hat{I}_d(s)=\hat{I}_r(s)-\hat{I}_l(s),\:
\hat{Q}_d(s)=\hat{I}_r(s)+\hat{I}_l(s)
\end{equation}
with the commutation rules
\begin{eqnarray}
&&[\hat{I}_l(s),\hat{I}_r(s')]=0,\\
&&[\hat{I}_l(s),\hat{I}_l(s')]=[\hat{I}_r(s'),\hat{I}_r(s)]=ie^2\partial_s(s-s')/2\pi.\nonumber
\end{eqnarray}
The last useful set of commutators is
\begin{eqnarray}
&&
[\hat{H}_{d},\hat{Q}_d(s)]=i\hbar\partial_s\hat{I}_d(s),\nonumber\\
&&
[\hat{H}_{d},\hat{I}_d(s)]=-i\hbar\partial_s\hat{Q}_d(s),\nonumber\\
&&
[\hat{H}_{d},\hat{I}_r(s)]=i\hbar\partial_s\hat{I}_r(s),\nonumber\\
&&
[\hat{H}_{d},\hat{I}_l(s)]=-i\hbar\partial_s\hat{I}_l(s).\label{comhd}
\end{eqnarray}
Similarly to (\ref{avg}),
equilibrium averages for the decoupled detector ($\lambda=0$) are
\ba
&&\mathrm{Tr}\hat{\rho}_d\hat{I}_d(\omega)=0,\nonumber\\
&&\mathrm{Tr}\hat{\rho}_d\hat{Q}_d(\omega)=2\pi n\delta(\omega),\nonumber\\
&&\mathrm{Tr}\hat{\rho}_d\hat{I}_l(\alpha)
\hat{I}_r(\beta)=0,\label{avgd}\\
&&\mathrm{Tr}\hat{\rho}_d\hat{I}_l(\alpha)\hat{I}_l(\beta)=
\mathrm{Tr}\hat{\rho}_d\hat{I}_r(\beta)\hat{I}_r(\alpha)=\nonumber\\
&&
=\frac{e^2}{2}\delta(\alpha+\beta)(w_d(\alpha)+\alpha),\nonumber
\ea
where $n$ is average charge density and
$w_d(\omega)=\omega\coth(\hbar\omega/k_BT_d)$. We shall put $n=0$
assuming that the average charge is screened out. 
In the Heisenberg picture, we have
\begin{eqnarray}
&&(\partial_t-\partial_s)\hat{I}_l(s,t)=-\partial_s\lambda(s)\hat{Q}(t),\nonumber\\
&&(\partial_t+\partial_s)\hat{I}_r(s,t)=\partial_s\lambda(s)\hat{Q}(t).\nonumber
\end{eqnarray}
The general solution is
\begin{eqnarray}
&&\hat{I}_l(s,t)=\hat{I}_l(t+s)-\partial_s\int_s^\infty ds'
\lambda(s')\hat{Q}(t+s-s'),\nonumber\\
&&\hat{I}_r(s,t)=\hat{I}_r(s-t)+\partial_s\int_{-\infty}^s ds'
\lambda(s')\hat{Q}(t-s+s').\nonumber
\end{eqnarray}
Hence, we have in the Heisenberg picture
\begin{eqnarray}
&&2\hat{\tilde I}(s,t)=2\partial_s\int ds'\lambda(s')\hat{Q}(t-|s-s'|)+\label{he1}\\
&&\hat{I}_r(s-t)-\hat{I}_r(-s-t)-\hat{I}_l(t+s)
+\hat{I}_l(t-s)\nonumber
\end{eqnarray}
and
\begin{eqnarray}
&&\int ds\;\lambda(s)\hat{Q}_d(s,t)=
\int ds\;\lambda(s)(\hat{I}_r(s-t)+\hat{I}_l(s+t))\nonumber\\
&&-2\int_0^\infty ds\int ds'\;\lambda(s')\lambda'(s-s')\hat{Q}(t-s).\label{he2}
\end{eqnarray}
To find $\mathcal S$, we need the 
Keldysh generating functional,\cite{kame}
\begin{eqnarray}
&&
e^{S[\chi,\phi]}=\mathrm{Tr}\hat{\rho}\times\label{sgen}\\
&&
\tilde{\mathcal T}e^{\int idt \left[\chi^-_Q(t)\hat{Q}(t)+\int ds(\chi_r^-(s,t)\hat{I}_r(s,t)
+\chi^-_l(s,t)\hat{I}_l(s,t))\right]}\times\nonumber\\
&&
\mathcal T e^{\int idt \left[\chi^+_Q(t)\hat{Q}(t)+\int ds(\chi_r^+(s,t)\hat{I}_r(s,t)
+\chi^+_l(s,t)\hat{I}_l(s,t))\right]}\nonumber
\end{eqnarray}
where $\chi^\pm=\chi/2\pm\phi$ and $S[0,\phi]=0$.
The functional is related to a quasiprobability in presence of external fields
\begin{eqnarray}
&&\varrho[I_r,I_l,Q;\phi]=\int D\chi\:e^{S[\chi,\phi]}\times\\
&&e^{-\int idt \left[\chi_Q(t)Q(t)+\int ds(\chi_r(s,t)I_r(s,t)
+\chi_l(s,t)I_l(s,t))\right]}\nonumber
\end{eqnarray}
One can show that Heisenberg equations (\ref{he1}) and (\ref{he2}) are satisfied for
the quasiprobability and
$I_r$, $I_l$ and $Q$ instead of the corresponding operators.
It can be also shown that the backaction of the detector on the system can be simplified
by the classical mapping $\hat{Q}\to Q$,
$\hat{Q}_d\to Q_d$ in Hamiltonian (\ref{hamdet}) used in (\ref{sgen}),
\begin{equation}
\hat{H}_I(t)=-2\pi\hbar\int ds\lambda(s)(Q_d(s,t)\hat{Q}+\hat{Q}_d(s,t)Q(t))/e^2.
\end{equation}
The final generating functional reads
\begin{eqnarray}
&&e^{\mathcal S[\chi]}
=\int D\phi D\xi D\eta\;e^{\int idt(\xi(t)\eta(t)-\chi(t)\phi(t)/2)}\times\nonumber\\
&&e^{-\int d\omega \left[|\chi(\omega)|^2w_d(\omega)/32\pi^2+|\phi(\omega)|^2/2w_d(\omega)\right]}\times\nonumber\\
&&e^{\tilde{S}[-\dot{\chi}\ast\lambda-\xi,-2\pi\phi\ast\lambda+4\pi\tilde\eta]}\label{sss}
\end{eqnarray}
with $\tilde{S}$ defined by (\ref{genf02}) and 
$
\tilde{\eta}(t)=\int_0^\infty ds\dot{\lambda}\ast\lambda(s)\eta(t-s)$,
 $w_d(\omega)=\omega\coth(\hbar\omega/k_BT_d)$.
In the case of many tapes, (\ref{weak2}) follows from
\begin{eqnarray}
&&e^{\mathcal S[\chi]}
=\int D\phi D\xi D\eta\;e^{\int idt(\xi(t)\eta(t)-\chi(t)\sum_c\phi_c(t)/2)}\times\nonumber\\
&&e^{-\int d\omega \left[|\chi(\omega)|^2\sum_cw_c(\omega)/32\pi^2+\sum_c|\phi_c(\omega)|^2/2w_c(\omega)\right]}\times\nonumber\\
&&e^{\tilde{S}[-\xi-\sum_c\dot{\chi}\ast\lambda_c,-2\pi\sum_c\phi_c\ast\lambda_c+4\pi\tilde\eta]}
\label{sss2}
\end{eqnarray}
with $\tilde{\eta}(t)=\sum_c\int_0^\infty ds\dot{\lambda}_c\ast\lambda_c(s)\eta(t-s)$.

\end{document}